\documentclass[journal=jacsat,manuscript=manuscript,layout=twocolumn]{achemso}
\DeclareUnicodeCharacter{2032}{\ensuremath{\prime}}
\DeclareUnicodeCharacter{0301}{\ensuremath{\prime}}

\usepackage[version=3]{mhchem} % Formula subscripts using \ce{}
\usepackage{multicol}
\SectionNumbersOn
\author{Karina Sogomonyan}
\affiliation[KU Leuven]
{KU Leuven, Department of Chemistry, Celestijnenlaan 200F, 3001 Leuven, Belgium.}
\email{karina.sogomonyan@kuleuven.be}
\author{Malek Ben Khalifa}
\affiliation[KU Leuven]
{KU Leuven, Department of Chemistry, Celestijnenlaan 200F, 3001 Leuven, Belgium.}
\author{Jérôme Loreau}
\affiliation[KU Leuven]
{KU Leuven, Department of Chemistry, Celestijnenlaan 200F, 3001 Leuven, Belgium.}
\email{jerome.loreau@kuleuven.be}

\title[An \textsf{achemso} demo]
  {Rotational Excitation of Vinyl Cyanide by Collisions with Helium Atoms at Low Temperature}

\keywords{vinyl cyanide, interaction potential, interstellar medium, scattering, rate coefficients}

\begin{document}

%%%%%%%%%%%%%%%%%%%%%%%%%%%%%%%%%%%%%%%%%%%%%%%%%%%%%%%%%%%%%%%%%%%%%
%% The "tocentry" environment can be used to create an entry for the
%% graphical table of contents. It is given here as some journals
%% require that it is printed as part of the abstract page. It will
%% be automatically moved as appropriate.
%%%%%%%%%%%%%%%%%%%%%%%%%%%%%%%%%%%%%%%%%%%%%%%%%%%%%%%%%%%%%%%%%%%%%
%\begin{tocentry}

%\end{tocentry}

%%%%%%%%%%%%%%%%%%%%%%%%%%%%%%%%%%%%%%%%%%%%%%%%%%%%%%%%%%%%%%%%%%%%%
%% The abstract environment will automatically gobble the contents
%% if an abstract is not used by the target journal.
%%%%%%%%%%%%%%%%%%%%%%%%%%%%%%%%%%%%%%%%%%%%%%%%%%%%%%%%%%%%%%%%%%%%%
\begin{abstract}
Among the numerous molecular systems found in the interstellar medium (ISM), vinyl cyanide is the first identified olephinic nitrile. While it has been observed in various sources, its detection in Sgr B2 is notable as the 2$_{11}$-2$_{12}$ rotational  transition exhibits maser features. This indicates that local thermodynamic equilibrium conditions are not fulfilled, and an accurate estimation of the molecular abundance in such conditions involves solving the statistical equilibrium equations taking into account the competition between the radiative and collisional processes. This in turn requires the knowledge of rotational excitation data for collisions with the most abundant species - He or H$_2$. In this paper the first three-dimensional CH$_2$CHCN - He potential energy surface is computed using explicitly correlated coupled-cluster theory [(CCSD(T)-F12] with a combination of two basis sets. Scattering calculations of the rotational (de-)excitation of CH$_2$CHCN induced by He atoms are performed with the quantum-mechanical close-coupling method in the low-energy regime. Rotational state-to-state cross sections derived from these calculations are used to compute the corresponding rate coefficients. The interaction potential exhibits a high anisotropy, with a global minimum of $-53.5$ cm$^{-1}$ and multiple local minima. Collisional cross sections are calculated for total energies up to 100 cm$^{-1}$. By thermally averaging the cross-sections, collisional rate coefficients are determined for temperatures up to 20 K. A propensity favouring the transitions with $\Delta k_a=0$ is observed.
\end{abstract}

%%%%%%%%%%%%%%%%%%%%%%%%%%%%%%%%%%%%%%%%%%%%%%%%%%%%%%%%%%%%%%%%%%%%%
%% Start the main part of the manuscript here.
%%%%%%%%%%%%%%%%%%%%%%%%%%%%%%%%%%%%%%%%%%%%%%%%%%%%%%%%%%%%%%%%%%%%%

\section{Introduction} \label{introduction}
Over the past decades, molecules of increasing complexity have been discovered in various astronomical environments, with a rapid acceleration in molecular detections in recent years. Among those, the interstellar complex organic molecules (small organic molecules with 6 atoms or more) often denoted as COMs comprise a considerable part \cite{mcguire20222021}. They have been observed in star-forming regions, as well as in diffuse and dense interstellar clouds. COMs possess varying complexity and chemical composition: both saturated and unsaturated COMs containing oxygen, nitrogen and/or sulfur have been detected. Such omnipresence suggests that their studies will lead to a better understanding of stellar evolution and formation of planetary systems \cite{herbst2009complex}. However, radioastronomical detections are hindered for molecular species with small or zero dipole moments \cite{yamamoto2017introduction}. Nitriles and other highly polar N-containing molecules are therefore among the most observed COM species via their rotational spectrum.\\ 
Vinyl cyanide (or acrylonitrile, CH$_2$CHCN) is the first olefinic nitrile detected in space \cite{gardner1975detection,matthews1983detection,agundez2008detection}. Nitriles are systems of prebiotic importance, considered among other C-N bond-bearing molecules to be potential precursors for aminoacids and nucleobases \cite{balucani2009elementary}. While vinyl cyanide itself is regarded as a candidate for cell-like membrane formation in anoxygenic environments \cite{stevenson2015membrane}, its olefinic nature also brings into light its terrestrial chemistry: it is prone to polymerisation and is very reactive overall. Vinyl cyanide was initially detected in space in Sgr B2 through its $2_{11}$-$2_{12}$ transition at 1.37 GHz, which displays maser features that indicate that local thermodynamic equilibrium conditions (LTE) are not fulfilled \cite{gardner1975detection}. Moreover, vinyl cyanide was also detected in the dense dark Taurus Molecular Cloud (TMC-1) \cite{matthews1983detection}. The authors report a column density of $3\times10^{12}$ cm$^{-2}$ using an excitation temperature $T_{ex}=5$ K  assuming LTE conditions. Such an assumption is made since the hyperfine components have intensity ratios matching the ones expected in LTE.  Furthermore, vinyl cyanide was observed in IRC +10216 with transitions involving levels up to  $j=11$ \cite{agundez2008detection}. Vinyl cyanide was also detected in Titan's atmosphere \cite{palmer2017alma} and in Orion KL in its ground and vibrationally excited states \cite{lopez2014laboratory}.\\
Chemical models utilizing large networks of elementary chemical and physical interstellar processes are employed as a means to improve our understanding of molecular evolution in space \cite{wakelam2010reaction}. Accordingly, a combined experimental and theoretical effort is needed to provide the data necessary for modelling. The formation of Vinyl cyanide was investigated experimentally via crossed beam approach \cite{balucani2000crossed}, and a possible formation path in the  interstellar medium is through a barrierless neutral-neutral reaction: C$_2$H$_4$($X^1A_g$) + CN($X^2\Sigma^+$). It was also reported that this reaction path is preferable to the one leading to the CH$_2$CHNC isomer. These results were later confirmed through a theoretical investigation by an RRKM-master equation analysis \cite{vereecken2003temperature}. Additionally, vinyl cyanide can be involved in further reactions with CN radicals to form more complex nitriles through similar barrierless fast reactions\cite{marchione2022unsaturated}.\\
Determining the abundances of molecular species in interstellar environments can be a challenging task as LTE conditions do not necessarily hold. Instead, one has to assume non-LTE excitation where the population of molecular levels is determined by the competition between  radiative and collisional processes. In this case, the accurate estimation of molecular abundances requires the knowledge of precise collisional excitation rate coefficients with the most abundant interstellar species (H$_2$ and He).
However, the collisional properties of the majority of COMs remain unknown. The increasing size of the systems presents computational challenges due to the strong anisotropy of the potential energy surface (PES) and the large number of rotational levels that can be populated at low energy. Collisional studies for COMs-H$_2$ are especially demanding \cite{dagdigian2024rotational,walker2017inelastic,khalifa2019interaction,BenKhalifa2024}, and He is often used as a proxy for para-H$_2$ ($j=0$), since both are spherical, while the mass difference is taken into account by scaling the rate coefficients. \cite{doi:10.1021/acs.jpca.2c06925,faure2019interaction,10.1093/mnras/stad3201,demes2024first}  \\
In this paper we present cross sections and rate coefficients for the collisional rotational (de-)excitation of the vinyl
cyanide with He atoms based on a newly developed
potential energy surface and fully quantum scattering calculations.
In section \ref{PES_section} we present the \textit{ab initio} study of the CH$_2$CHCN - He system and the construction of the underlying PES, obtained with the coupled cluster method using a combination of two basis sets. In section \ref{dynamics_section} we report the quantum dynamics study of these collisions and discuss the resulting inelastic scattering cross sections at low energy. 
The corresponding (de-)excitation rate coefficients are derived for kinetic temperatures up to 20 K based on the quantum scattering calculations and presented in section \ref{rates_section}. Conclusions and outlooks are discussed in section \ref{Conclusions_section}.

\section{Potential energy surface} \label{PES_section}
\subsection{Geometry}

\begin{figure}[h]
    \centering
    \includegraphics[width=0.6\linewidth]{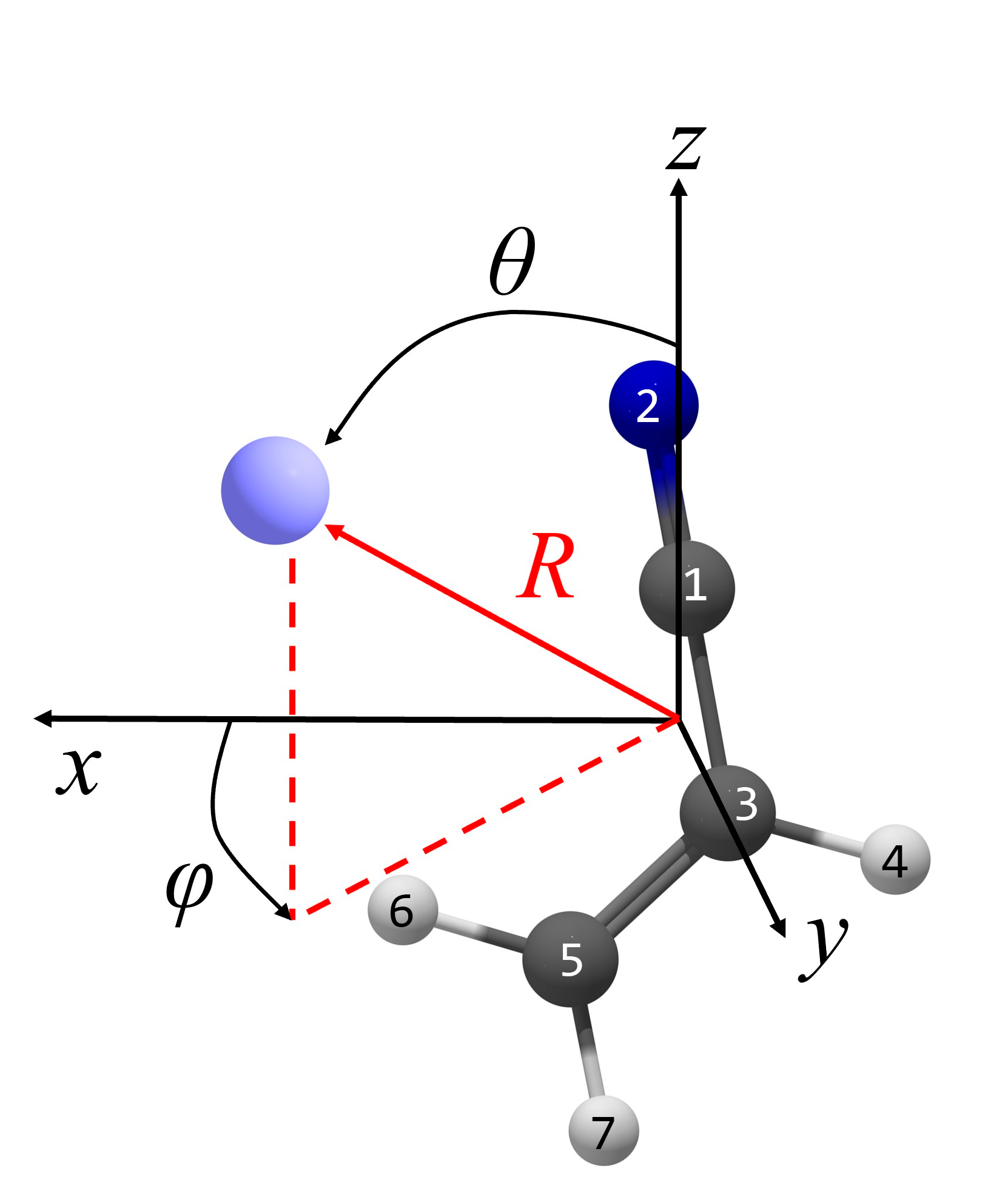}
    \caption{Jacobi coordinate system for the CH$_2$CHCN-He complex. The coordinate system origin is in the centre of mass of CH$_2$CHCN. The molecule lies in the $xz$ plane.}
    \label{fig:coordsys}
\end{figure}

Our first aim is to investigate the interaction between the asymmetric top molecule CH$_2$CHCN and a helium atom in their respective ground electronic states. The lowest vibrational mode for vinyl cyanide is $v=242$ cm$^{-1}$ \cite{motte1995electronic}. For high-energy collisions the first vibrational mode should be taken into account while constructing the PES, whereas for low-temperature environments of ISM, a rigid rotor approximation can be applied.\\
The geometry parameters determined from the experimental rotational spectra of vinyl cyanide \cite{buck1976structure} were used as internal coordinates: $r$(C$_1$$\equiv$N$_2$) = 1.164 \AA, $r$(C$_1$--C$_3$) = 1.426 \AA, $r$(C$_3$=C$_5$) = 1.339 \AA, $r$(C$_5$--H$_6$) = 1.088 \AA, $r$(C$_5$--H$_7$) = 1.088 \AA, $r$(C$_3$--H$_4$) = 1.086 \AA, $\angle$(C$_1$C$_3$C$_5$) = 122.6$^{\circ}$, $\angle$(H$_4$C$_3$C$_5$) = 121.7$^{\circ}$,$\angle$(C$_3$C$_1$N$_2$) = 180$^{\circ}$. The coordinate system used for the CH$_2$CHCN interaction with helium is presented in Figure \ref{fig:coordsys}. The origin of the axes is the mass centre of vinyl cyanide, while the vinyl cyanide molecule is oriented in its principal axis frame. The molecule is planar and lies in the $xz$ plane. The interaction with helium is defined by three intermolecular coordinates: $R$, $\theta$, $\phi$. $R$ is the magnitude of the vector $\mathbf{R}$ that connects the centre of mass of the vinyl cyanide molecule and the helium atom, it represents the distance between the two colliding partners. $\theta$ is the polar angle between the vector $\mathbf{R}$ and the axis $z$, aligned with the principal inertia axis $a$ of vinyl cyanide, while $\phi$ is the azimuthal angle. 

\subsection{Computational methods}
\begin{figure}
    \centering
    \includegraphics[width=0.9\linewidth]{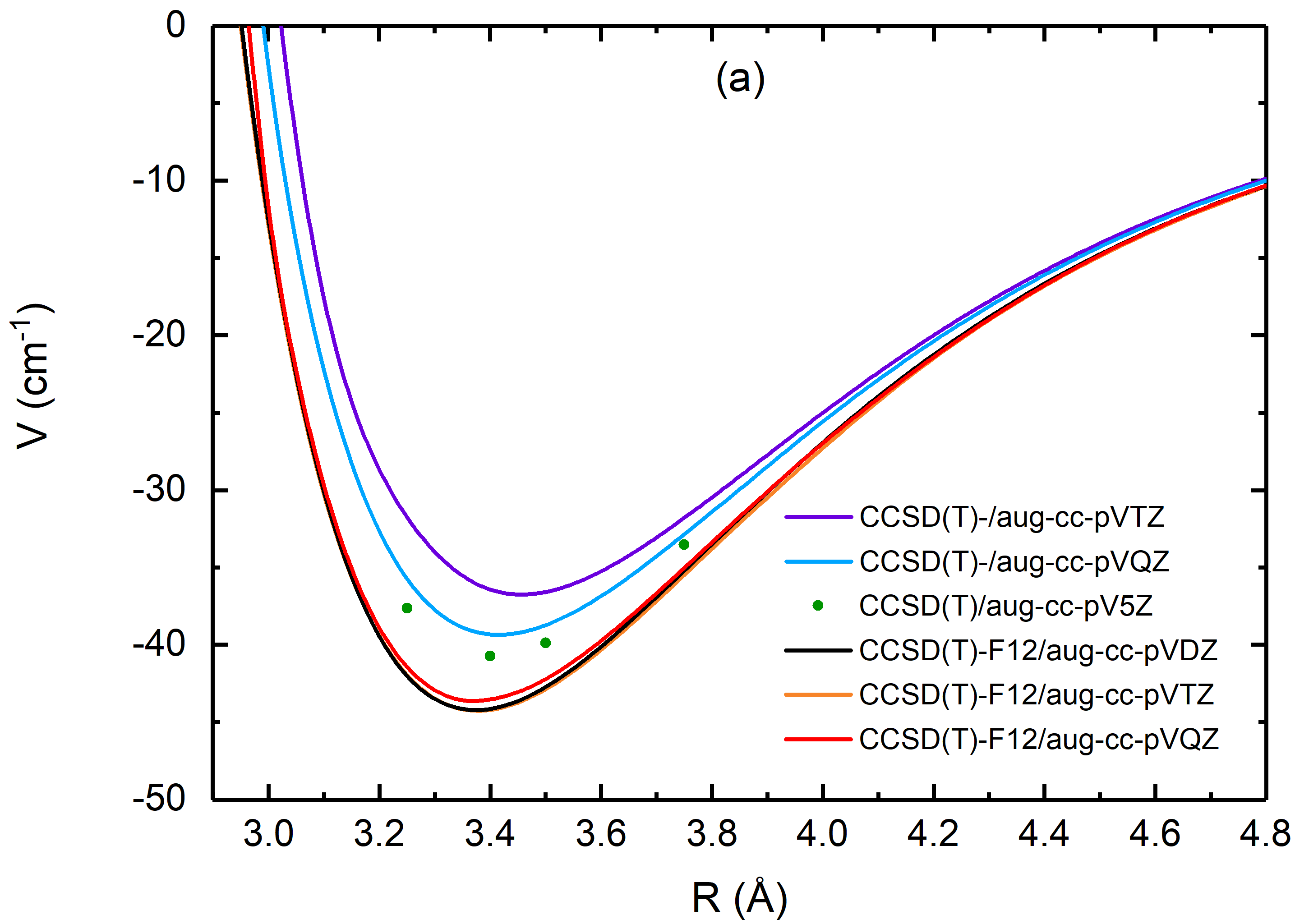}
    \includegraphics[width=0.9\linewidth]{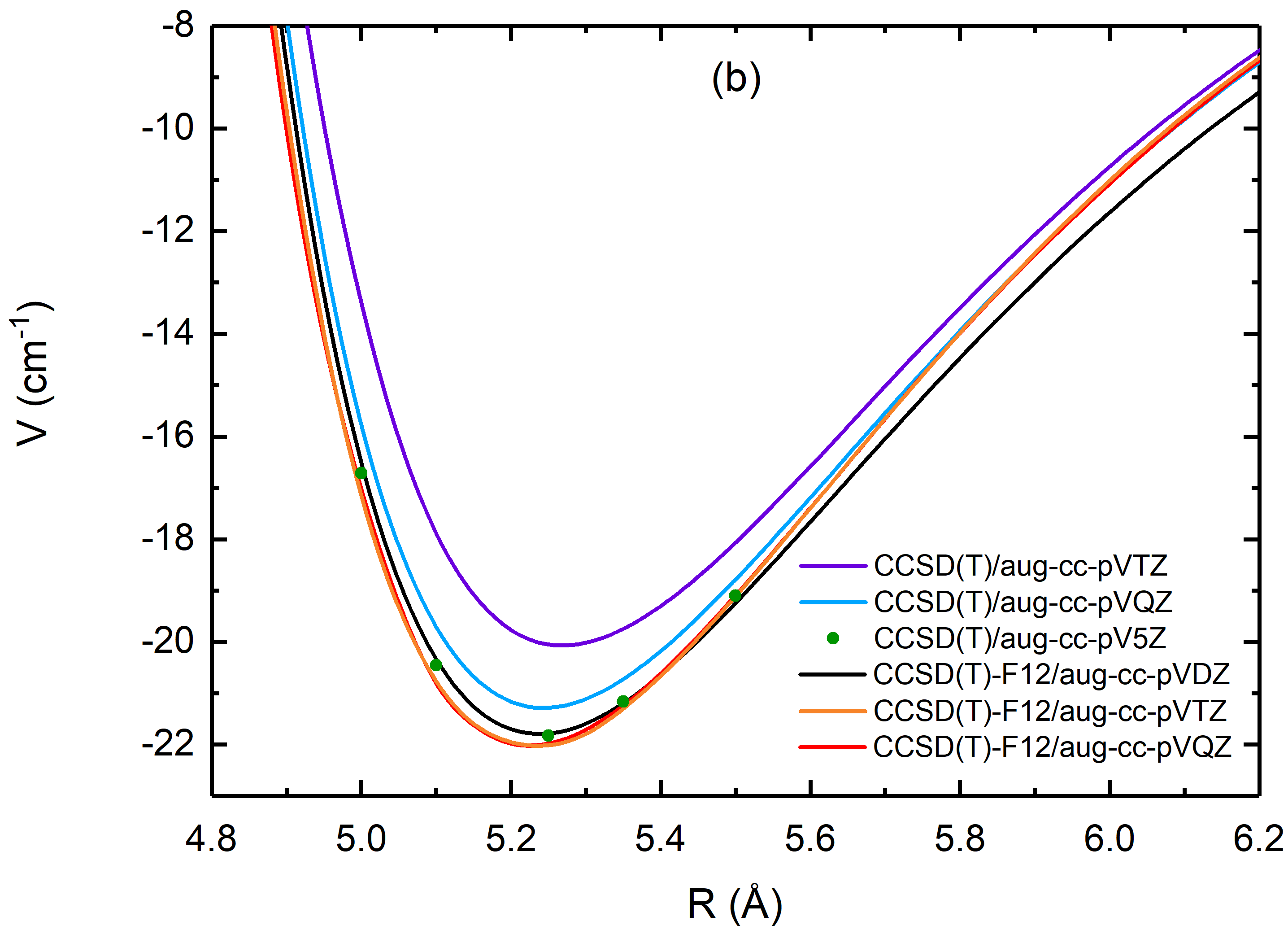}
    \caption{Potential energy cuts (cm$^{-1}$) for the CH$_2$CHCN - He complex calculated at different levels of theory as a function of $R$ for $\phi$ = 90$^{\circ}$, $\theta$ = 90$^{\circ}$ (panel a) and $\phi$ = 0$^{\circ}$, $\theta$ = 16.11$^{\circ}$ (panel b).}
    \label{fig:benchmark}
\end{figure}

The \textit{ab initio} calculations were carried out with the MOLPRO (version 2020.2) program package \cite{10.1063/5.0005081}. The counterpoise correction procedure of Boys and Bernardi \cite{boys1970calculation} was taken into account to compute the \textit{ab initio} interaction energy:
\begin{equation}\label{eq:1}
    V(R,\theta,\phi)=V_{\rm{Mol-He}}(R,\theta,\phi)-V_{\rm{Mol}}(R,\theta,\phi)-V_{\rm{He}}(R,\theta,\phi)
\end{equation}
To find a suitable method for the construction of the full PES, a set of test potential energy surface cuts was calculated. For several fixed orientations of the He atom with respect to the vinyl cyanide molecule, we performed a comparison using a standard CCSD(T) \cite{DEEGAN1994321} and explicitly correlated CCSD(T)-F12a method with single, double, and perturbative triple excitations \cite{adler2007simple}. Here we used the augmented correlation-consistent basis sets developed by Dunning et al. \cite{10.1063/1.456153}. For CCSD(T) calculations the aug-cc-pVnZ (aVnZ for short), n = T,Q,5 basis sets were used, while for CCSD(T)-F12a we limited calculations to aVnZ, n=D,T basis sets. Figure \ref{fig:benchmark} displays potential energy cuts for two geometries from the chosen set of orientations. These plots demonstrate the usual behaviour for the CCSD(T)/aVnZ series: the correlation-consistent basis sets lead to a deeper potential well for the increasing basis size and converge toward a basis set limit. On the other hand, the CCSD(T)-F12a/aVnZ series provides near identical results for all three tested basis sets (n=D, T, Q) but does not demonstrate consequent convergence. In some cases (panel a) it also overestimates the correlation energies. This is a known behaviour associated with these explicitly correlated methods \cite{adler2007simple}. Considering that CCSD(T)-F12a/aVnZ, n = D,T calculations provide the closest results to the CCSD(T)/aug-cc-pV5Z values and that the results obtained with these two methods are almost the same, a combination of the two was chosen to perform the full PES calculations, as detailed below. This will provide an accurate PES while considerably reducing the CPU times and disk space.

\subsection{Methodology}
\label{methodology}
The construction of a precise potential energy surface consisted of several parts: \textit{ab initio} calculations of the interaction energies with CCSD(T)-F12a/aVnZ, n = D,T methods, their respective fitting, and the merging of the two analytical fits to obtain the radial coefficients used in scattering calculations.
Quantum scattering calculations require an expansion of the angular dependence of the PES over spherical harmonics. In terms of the spherical coordinates ($R$,  $\theta$, $\phi$) this expansion can be written as:
\begin{equation}
V(R,\theta,\phi)=\sum_{l=0}^{l_{\max}}\sum_{m=-l}^{l}v_{lm}(R)Y_{l}^{m}(\theta,\phi)\label{eq:expansion}
\end{equation}
or alternatively, considering the  properties of the spherical harmonics, as:
 \begin{equation}\label{eq_fit}
V(R,\theta,\phi)=\sum_{l=0}^{l_{\max}}\sum_{m=0}^{l}v_{lm}(R)\frac{Y_{l}^{m}(\theta,\phi)+(-1)^{m}Y_{l}^{-m}(\theta,\phi)}{1+\delta_{m,0}}
\end{equation}
where $v_{lm}(R)$ and $Y_{l}^{m}(\theta,\phi)$ are the radial coefficients to be computed and the normalised spherical harmonics respectively, and $\delta_{m,0}$ is the Kronecker symbol. From the PES parametrised by $R$,  $\theta$, and $\phi$, we obtain  continuous radial coefficients $v_{lm}(R)$ through a least-squares fit for each value of the integers $l$ and $m$. For every analytical fitting procedure, we  adjust the parameters $l_{\max}$ and $m_{\max}$ to ensure an accurate representation of the potential while limiting the CPU time increase in the scattering calculations.\\
The full procedure to the PES was as follows. 
In a first step, we compute a dense grid of \textit{ab initio} points with the lower accuracy method (CCSD(T)-F12a/aVDZ). We then perform the analytical fitting procedure, leading to the radial coefficients denoted as $v_{lm}^{\textrm{DZ}}$. The analytical potential that can be reconstructed from the $v_{lm}^{\textrm{DZ}}$ coefficients is denoted as $V^{\textrm{DZ}}$.
In a second step, we compute \textit{ab initio} energies with a higher accuracy method (CCSD(T)-F12a/aVTZ) on a sparser grid and we perform the analytical fitting procedure for this reduced PES. Since the number of angles considered will be smaller, the parameters $l_{\max}$ and $m_{\max}$ in Eq. (\ref{eq_fit}) are smaller than for $V^{\textrm{DZ}}$. The resulting radial coefficients are denoted as  $v_{lm}^{\textrm{TZ,red}}$ and the reconstructed analytical potential is denoted as $V^{\textrm{TZ,red}}$.
We then obtain a fit of the DZ energies on the same coarser grid and with the same expansion parameters as for the $V^{\textrm{TZ,red}}$ potential. The resulting radial coefficients are denoted as $v_{lm}^{\textrm{DZ,red}}$ and the reconstructed potential as $V^{\textrm{DZ,red}}$.
In a third step, we calculate the difference of the radial coefficients with matching $l$ and $m$ parameters for every $R$: $v_{lm}^{\textrm{TZ,red}}-v_{lm}^{\textrm{DZ,red}}=v_{lm}^{\textrm{diff}}$. We can then construct the analytical potential $V^{\textrm{diff}}$ from the coefficients $v_{lm}^{\textrm{diff}}$ and add this potential to the $V^{\textrm{DZ}}$ potential, $V^{\textrm{DZ}}+V^{\textrm{diff}}=V^{\textrm{sum}}$. This potential is supposed to match the quality of the full PES performed with CCSD(T)-F12a/aug-cc-pVTZ.
Finally, $V^{\textrm{sum}}$ can be expanded with Eq.(\ref{eq_fit}) to obtain the resulting $v_{lm}^{\textrm{sum}}$ radial coefficients that are used for the scattering calculations.
\subsection{\textit{Ab initio} calculations}
$V^{\textrm{DZ}}$ was constructed from 50,616 \textit{ab initio} energies. The geometries were chosen for 72 intermolecular distances from 2.0 to 100.0 \AA, 37 values of $\theta$ ranging between [0-180$^{\circ}$] by steps of 5$^{\circ}$ and 19 values of $\phi$ in the range [0-180$^{\circ}$] by steps of 10$^{\circ}$. This large number of points  was found to be necessary to obtain an accurate analytical potential $V^{\textrm{DZ}}$ due to the strong anisotropy. 
$V^{\textrm{TZ,red}}$ was constructed from 8,029 \textit{ab initio} points in total. These were chosen for 31 intermolecular distances from 2.0 to 100.0 \AA, 37 values of $\theta$ ranging between [0-180$^{\circ}$] by steps of 5$^{\circ}$ and 7 values of $\phi$ in the range [0-180$^{\circ}$] by steps of 30$^{\circ}$. A dense grid in $\theta$ was chosen because preliminary tests of the fitting algorithm showed that sparser grids systematically led to spurious features in the fitted potential. This reflects the large anisotropy in $\theta$, further discussed in the next sections. \\
Due to the inclusion of perturbative triple excitations, the explicitly-correlated CCSD(T)-F12 methods are not size consistent \cite{knizia2009simplified}. For every fixed angular orientation the asymptotic value of the potential was subtracted from the interaction energies.

\subsection{Analytical fits}
$V^{\textrm{DZ}}$ is constructed from a grid containing 37 values of $\theta$ and 19 values of $\phi$, from which we were able to obtain radial coefficients up to $l_{\max}=16$ $m_{\max}=15$, resulting in 152 expansion terms. $V^{\textrm{TZ,red}}$ is obtained on a grid containing 37 values of $\theta$ and 7 values of  $\phi$, from which radial coefficients up to $l_{\max}=16$ $m_{\max}=6$ were calculated,  resulting in 98 expansion terms.
The reduced reconstructed potential $V^{\textrm{DZ,red}}$ was obtained through fitting of the aVDZ energies over the same grid as for $V^{\textrm{TZ,red}}$. The final expansion of the potential $V^{\textrm{sum}}$ was performed with the parameters $l_{\max}=16$, $m_{\max}=15$. To assess the quality of the obtained potential we performed an additional quality test: a set of 75 randomized $\textit{ab initio}$ points were calculated at the CCSD(T)-F12/aVTZ level of theory. The resulting accuracy of the fitting procedure was better than 1 cm$^{-1}$ for the attractive part of the PES.\\
The calculation of molecule-helium interaction potentials presents a complication of importance in the framework of collisional studies: the long-range part of the potential ($R\ge8$ \AA) usually demonstrates oscillations with an amplitude of up to 0.1 cm$^{-1}$. This known issue was also present in our case and needs to be addressed since the long-range region of the PES is crucial for the scattering calculations at low collisional energies. 
The functions $v_{lm}^{\textrm{sum}}$ obtained as a result of the steps described in section \ref{methodology} in the asymptotic region were adjusted by fitting 5 points in the $R$-grid at 11, 12, 13, 15, and 17 \AA\ to an expansion in inverse powers of $R$:
\begin{equation}
    v_{lm}^{\textrm{sum}}(R)=\sum_{n}c_{lmn}R^{-n}\label{longrange}
\end{equation}

\begin{figure}
    \centering
    \includegraphics[width=1\linewidth]{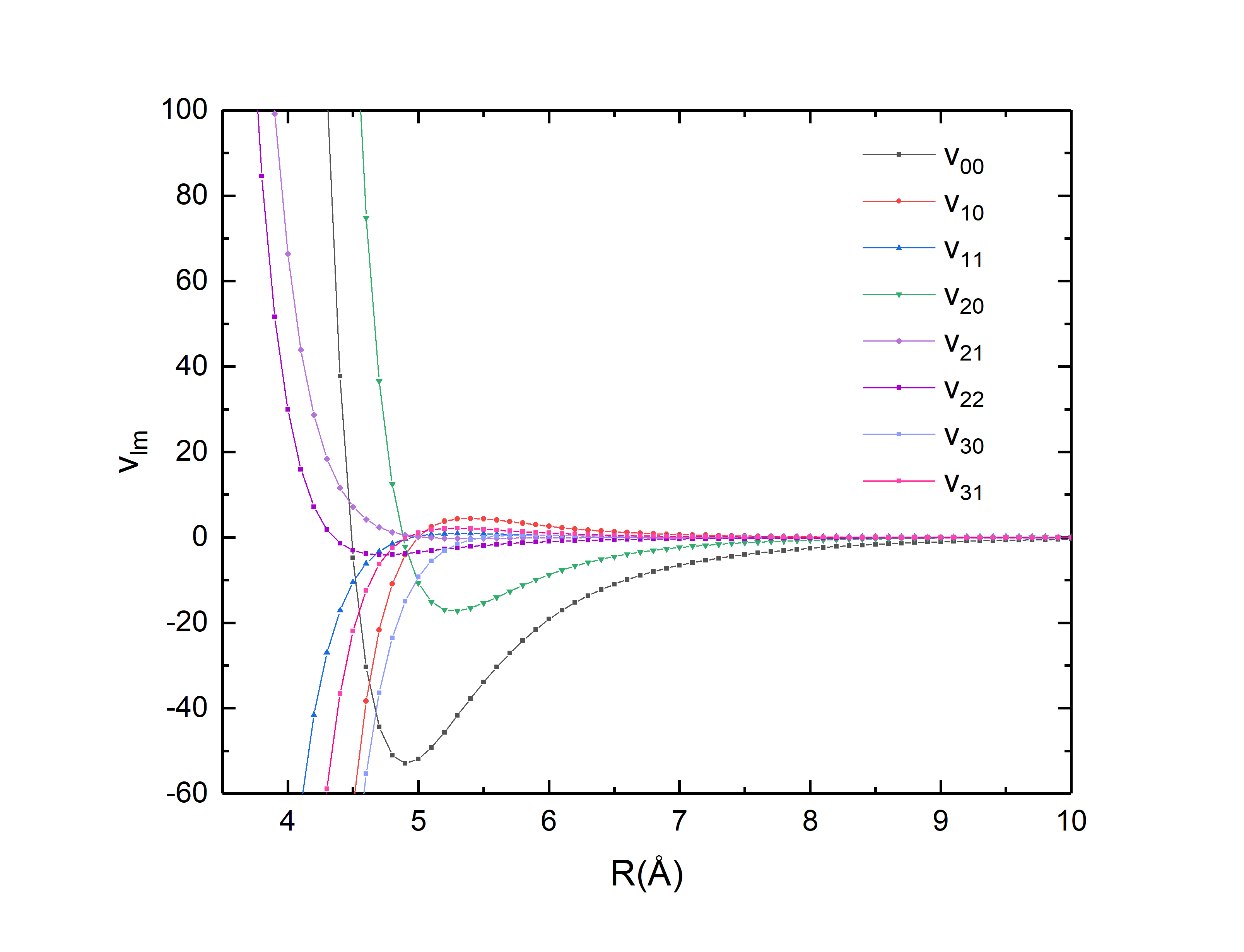}
    \caption{Dependence on $R$ of the first $v_{lm}$($R$) components for CH$_2$CHCN -- He with $l\leq3$.}
    \label{vlmdiff-label}
\end{figure}

We kept only the first two terms in the inverse $R$-power expansion, $n=n_i$ and $n=n_i+2$. The value of $n_i$ depends \cite{van1980ab} on the value of $l$: $n_i=6$ for $l=0$ and $2$, $n_i=7$ for $l=1$ and $3$, and $n_i=l+4$ for $l\ge4$. Each function $v_{lm}^{\textrm{sum}}$ is represented by two coefficients $c_{lmn}$. The analytic form (\ref{longrange}) for the asymptotic $v_{lm}^{\textrm{sum}}$ functions was used for the first 78 radial functions, corresponding to all couples ($l,m$) with $l<12$. All values of $v_{lm}^{\textrm{sum}}$ for $l\ge12$ and $R\ge8$ \AA\ were set to 0 in the long-range. Figure \ref{vlmdiff-label} shows the first 8 $v_{lm}^{\textrm{sum}}$ components.\\
Besides the isotropic term $v_{00}$ with a well depth of 53 cm$^{-1}$, for $m=0$ the term with $l=2$ slightly outweighs $v_{10}$, $v_{30}$, and the other anisotropic terms. This term is responsible for rotational transitions with $\Delta j=2$, which should affect the propensity rules of the rotational excitation.

\subsection{Description of the potential energy surface}
\begin{figure*} 
    \centering
    \includegraphics[width=.4\linewidth]{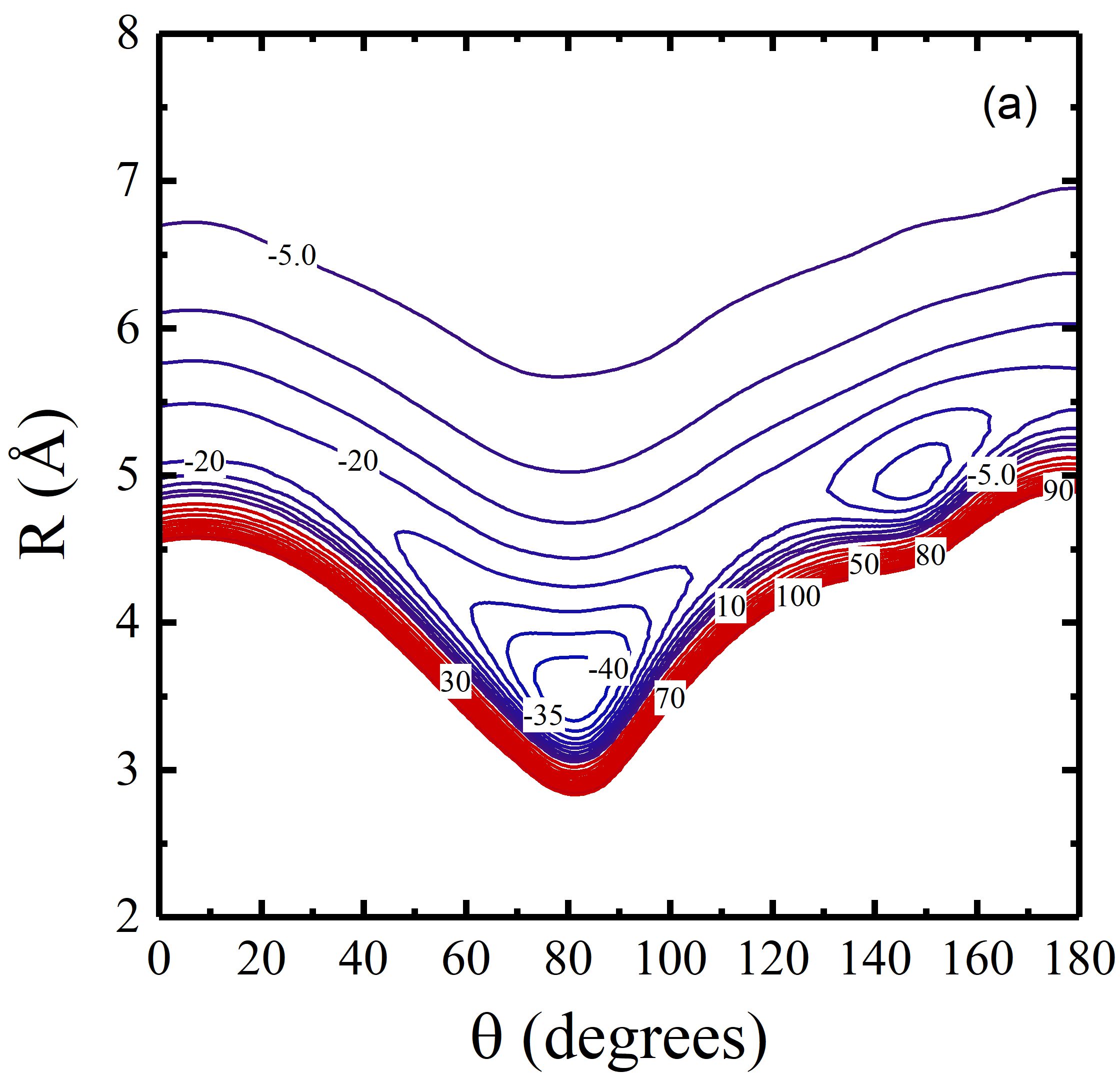}
    \includegraphics[width=.4\linewidth]{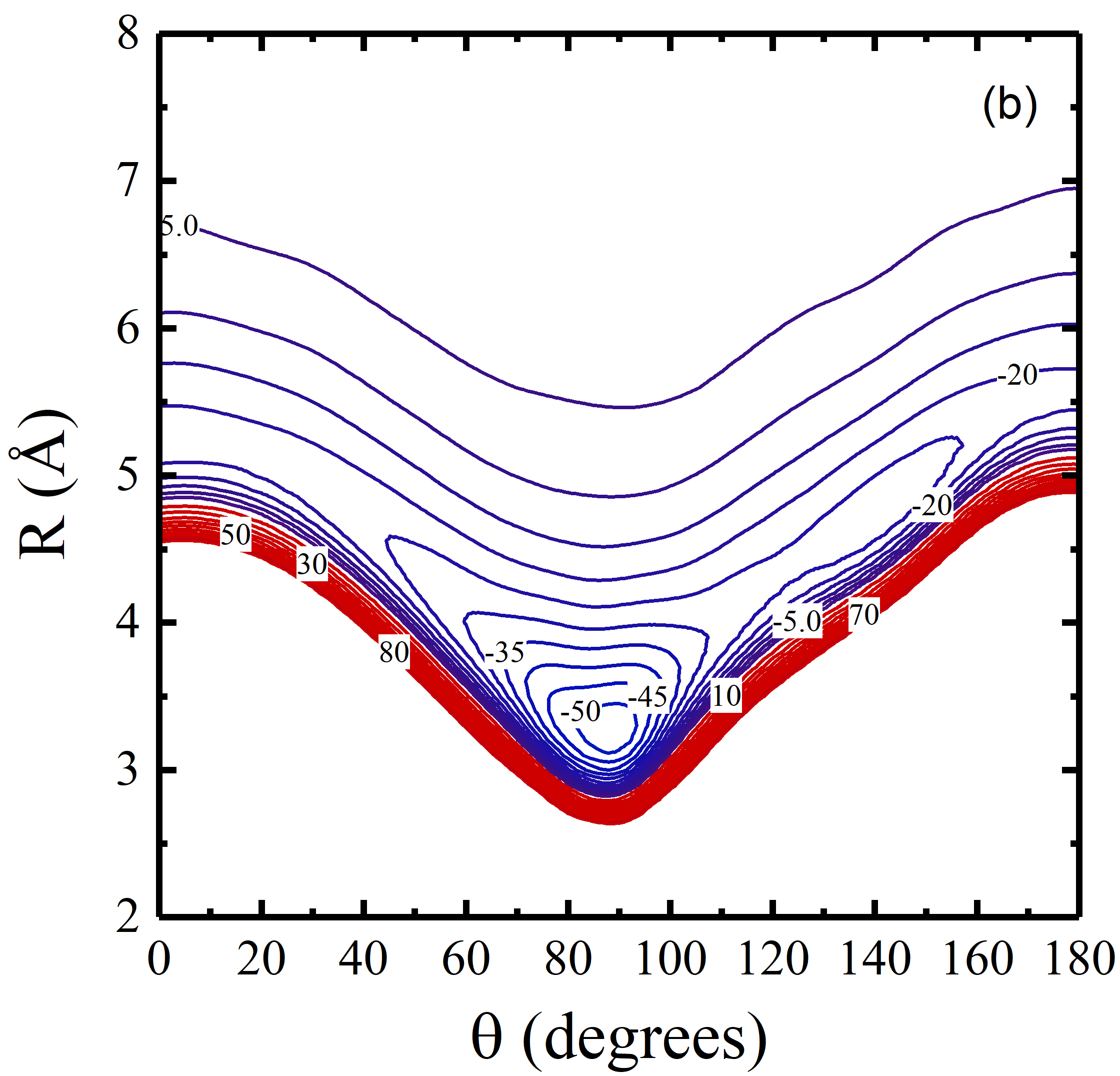}
    \includegraphics[width=.4\linewidth]{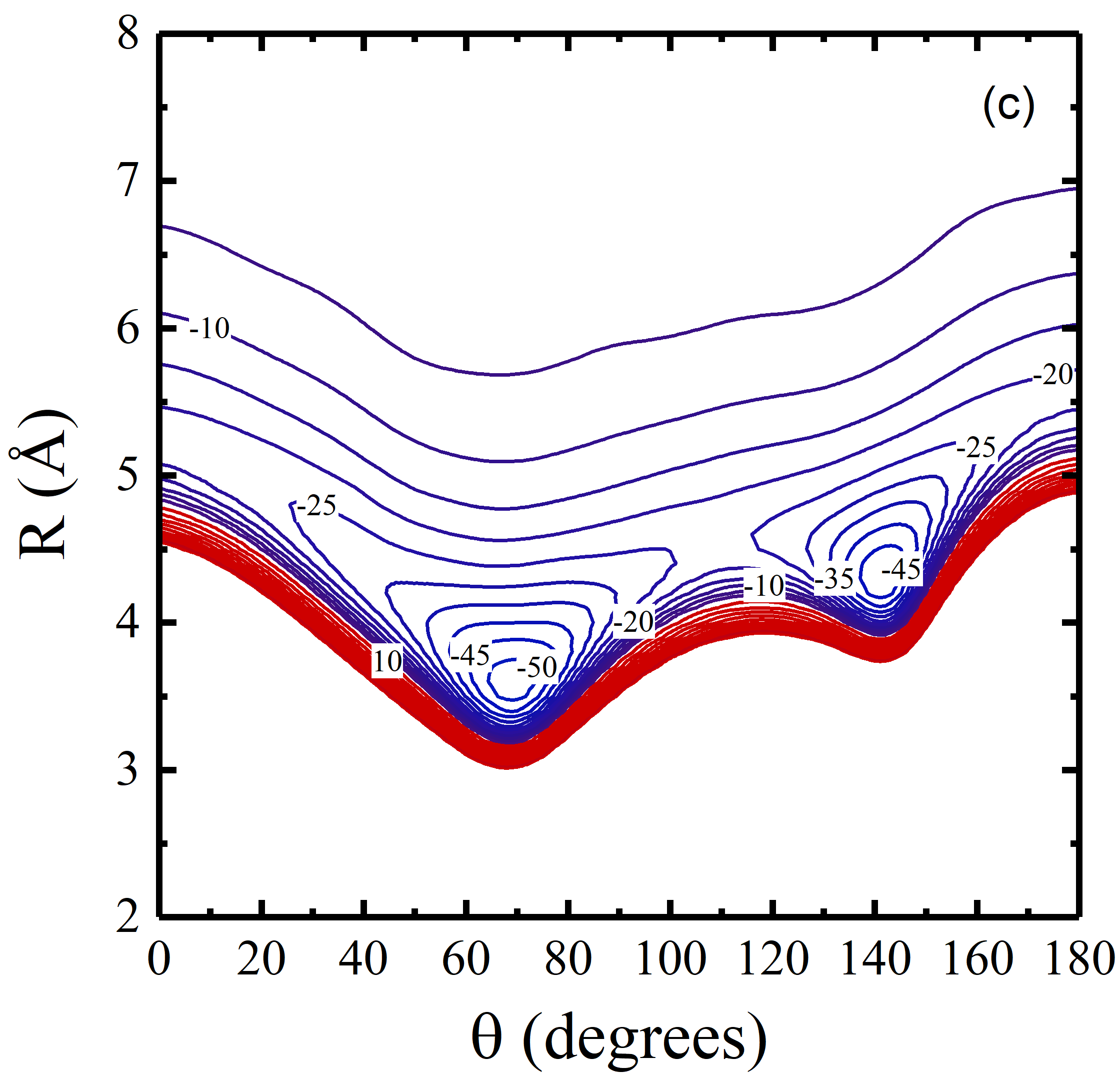}
    \includegraphics[width=.41\linewidth]{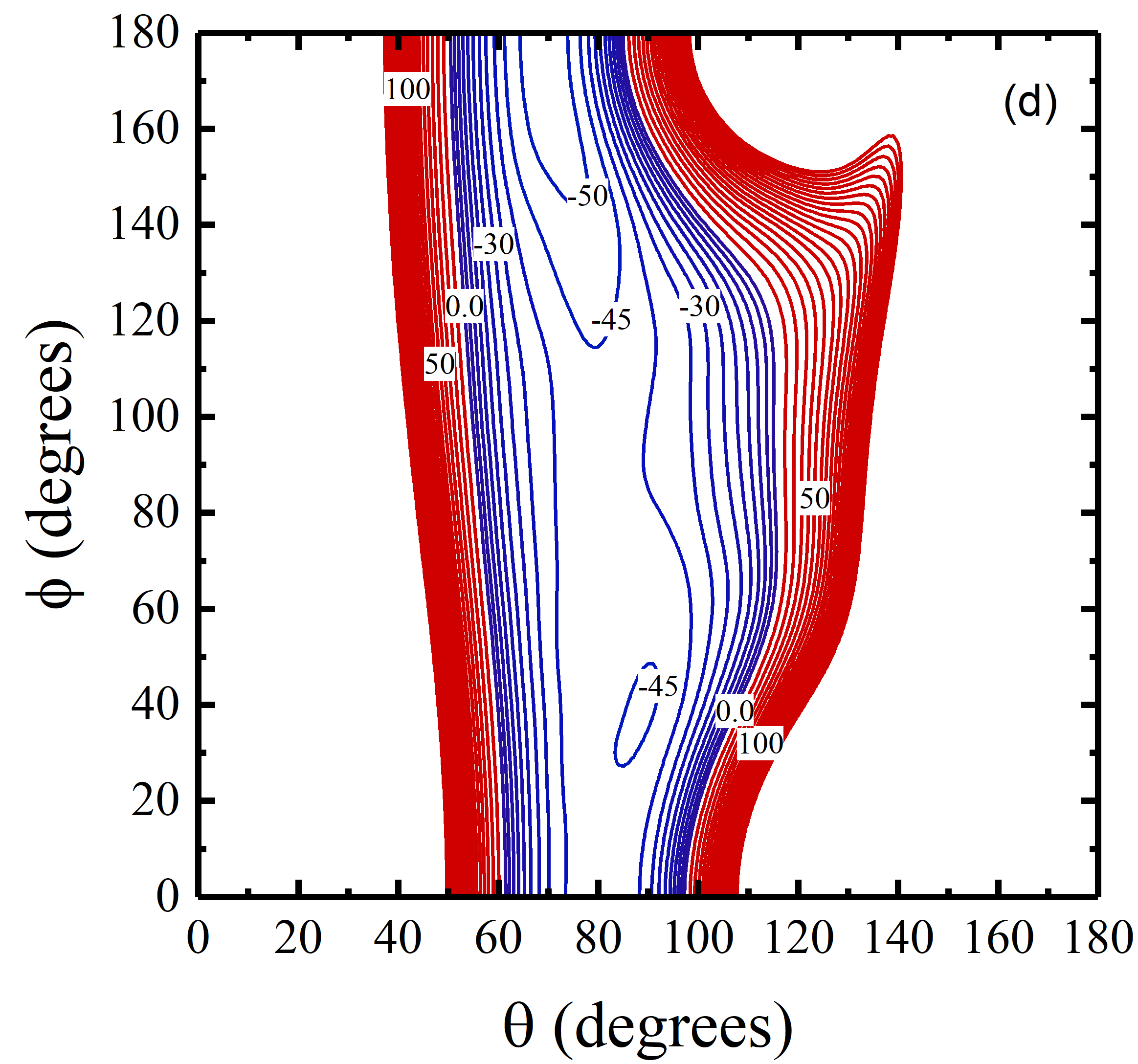}
    \caption{Two-dimensional contour plots of the interaction potential of the CH$_2$CHCN-He complex. Panels (a), (b) and (c) depict the PES as a function of $\theta$ and $R$ for $\phi=0^\circ$ (a), 51$^\circ$ (b) and 180$^\circ$ (c). Panel (d) shows PES as a function of $\phi$ and $\theta$ at $R=3.6$ \AA.}
    \label{fig:pes}
\end{figure*}

Figure \ref{fig:pes} displays two-dimensional contour plots of the CH$_2$CHCN-He van der Waals complex as a function of the two Jacobi coordinates $R$ and $\theta$ for $\phi=0^{\circ}$ (Panel a), $\phi=51^{\circ}$ (Panel b) and $\phi=180^{\circ}$ (Panel c), while panel (d) shows the potential as function of $\theta$ and $\phi$ for fixed $R$. Panels a and c correspond to the helium atom rotating around the molecule in the molecular plane. 

The global minimum of the PES is found at for $\phi=51^{\circ}$, $\theta=89^{\circ}$, $R=3.3$ \AA\, corresponding to the He atom hovering over the molecular plane and with a well depth of $V=-53.5$ cm$^{-1}$, as seen in panel (b). 

When the He atom is in the molecular plane, several minima are observed. 
For $\phi=0^{\circ}$ (panel (a)), a local minimum is found at $\theta=82^{\circ}$, $R=3.5$ \AA, with a well depth of $V= -44.6$ cm$^{-1}$. Another local minimum of $V=-32.7$ cm$^{-1}$ can be seen at $\theta=148^{\circ}$ and a much larger distance, $R=5.0$ \AA. The position of these minima correspond to the helium atom approaching between the hydrogen atom and the C$\equiv$N group attached to the different C-atoms of the vinyl group, and between two hydrogen atoms of the CH$_2$ group. The barrier between the minima is around 22 cm$^{-1}$. 
 For $\phi=180^{\circ}$ (panel (c)) we find a minimum with a well depth of $V=-52.9$ cm$^{-1}$ at $\theta = 69^{\circ}$, $R=3.6$ \AA\, while  another local minimum is located at $\theta = 141^{\circ}$, $R=4.3$ \AA\ with a well depth of $V=-49.2$ cm$^{-1}$. These minima correspond to the helium atom approaching between the hydrogen atom and the C$\equiv$N group attached to the same C-atom, and between two hydrogen atoms around the double bond of the vinyl group. The two local minima are separated by a barrier of 28.5 cm$^{-1}$. \\
Interestingly, when the He atom is located in the molecular plane the local minimum only differs from the global minimum by 0.6 cm$^{-1}$. The energy of the minimum of the PES for fixed $\phi$ decreases from $-44.6$ cm$^{-1}$ for $\phi=0$ to $-53.5$ cm$^{-1}$ for $\phi=51^\circ$ before increasing to $-47.1$ cm$^{-1}$ for $\phi=90^\circ$. It then decreases again until it reaches $-52.9$ cm$^{-1}$ for $\phi=180^\circ$.
 This behaviour is displayed in panel (d) of Figure \ref{fig:pes}, which illustrates a two-dimensional cut of the PES along the angular coordinates $\theta$ and $\phi$ with $R$ fixed to  $R=3.6$ \AA. We note that the energies of the minima in this figure are different from the ones discussed earlier, as the minima do not all occur for the same value of $R$.
 From this figure it is also clear that the anisotropy of the surface is much larger for rotations along the $\theta$ coordinate than the $\phi$ coordinate. 
 This indicates that the rotation of the He atom around the molecule along the $\phi$ coordinate would be nearly unhindered in the van der Waals complex CH$_2$CHCN-He.
\section{Dynamics}\label{dynamics_section}
\subsection{Rotational Hamiltonian}
\begin{figure}
    \centering
    \includegraphics[width=0.85\linewidth]{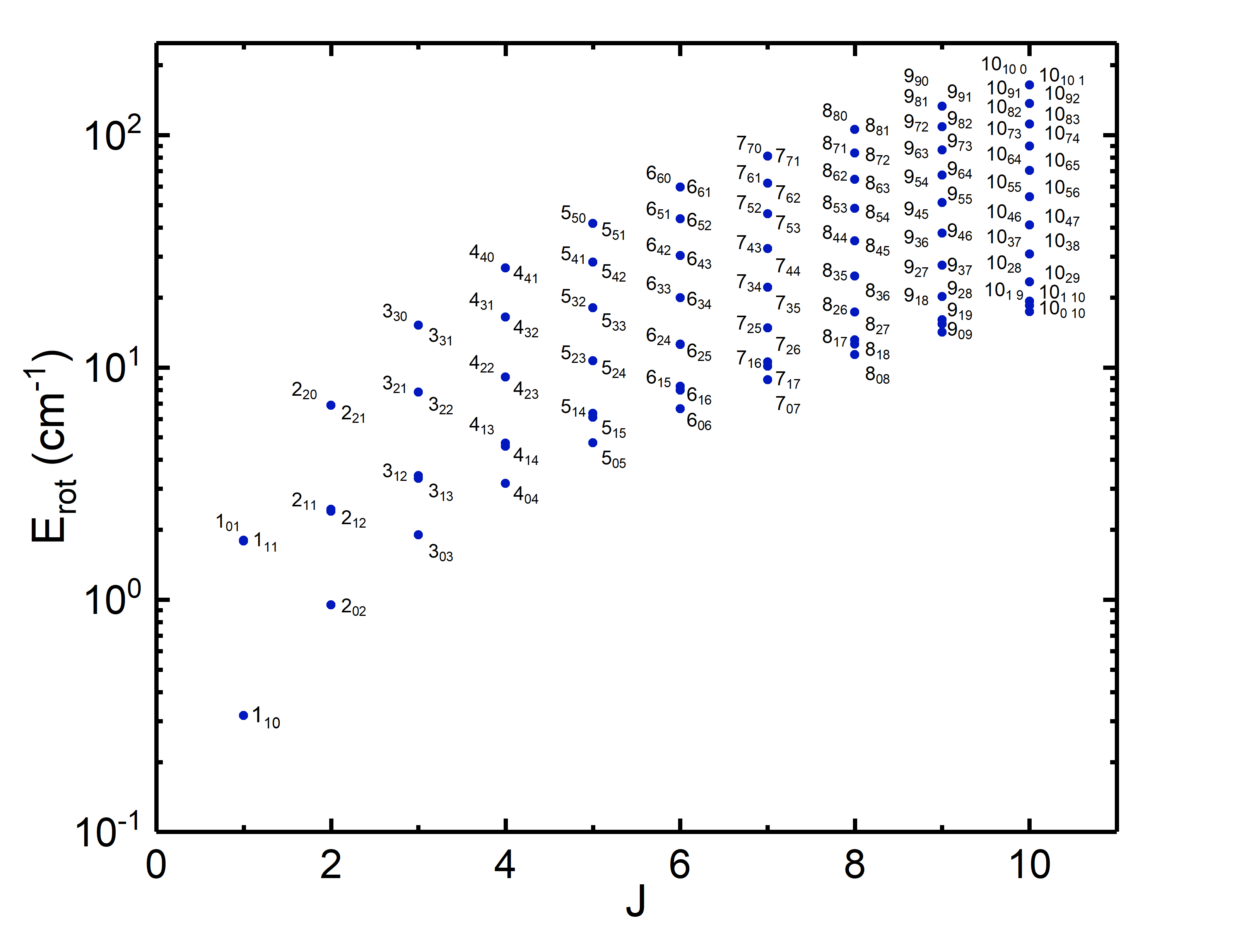}
    \caption{Rotational levels of CH$_2$CHCN up to $j = 10$.}
    \label{fig:rotlvl}
\end{figure}

Vinyl cyanide is an asymmetric top molecule with a rotational Hamiltonian written as \cite{schmitt2018structures}:
\begin{equation}\label{Hrot}
    H_{rot}=Aj_x^2+Bj_y^2+Cj_z^2-D_jj^4-D_{jk}j^2j_z^2-D_kj_z^4
\end{equation}
where $A$, $B$ and $C$ are rotational constants, $D_j$, $D_{jk}$, $D_k$ are first order centrifugal distortion constants of vinyl cyanide and $j_x$, $j_y$, $j_z$ are the projections of the angular momentum $j$ along the principal inertia axes: $j^2=j^2_x+j^2_y+j^2_z$. The rotational wave functions of a symmetric top molecule $\vert jkm \rangle $ are no longer eigenfunctions of the rotational Hamiltonian in equation (\ref{Hrot}), but they are used in a linear combination to express the asymmetric top wave functions $\vert j\tau m \rangle $ characterised by three quantum numbers $j$, $\tau$, $m$:
\begin{equation}\label{wavef}
    \vert j\tau m \rangle = \sum_k a_{\tau k}\vert jkm\rangle
\end{equation}
where $k$ denotes the projection of $j$ along the $z$-axis of the body-fixed frame, $m$ is the projection on the space-fixed $Z$-axis and $\tau$ is an integer $-j\leq\tau\leq j$ which orders the energy levels for a given value of $j$ \cite{flower2007molecular}.\\
The rotational levels of asymmetric top molecules are labeled by $k_a$ and $k_c$ numbers representing the projections of the quantum number $k$ along the axes of symmetry in the prolate and oblate symmetric top limits, respectively. The relation between $k_a$ and $k_c$ can also be expressed through $\tau$: $\tau$ = $k_a - k_c$. The rotational constants $A=1.63701$, $B=0.16583$, $C=0.15057$ (all in cm$^{-1}$) were taken from the Computational Chemistry Comparison and Benchmark Database (CCCBDB) \cite{147901}. The centrifugal distortion constants were taken from the literature \cite{lopez2014laboratory}: $D_j = 7.2797 \times 10^{-8}$, $D_{jk} = 2.8358 \times 10^{-6}$ and $D_k = 9.0559 \times 10^{-5}$ (all in cm$^{-1}$). The energy level diagram of vinyl cyanide is shown in Fig. \ref{fig:rotlvl}. 

\subsection{Scattering cross sections}
\label{scatcalc-subsect}
Considering the complex rotational structure of the system and the low temperatures of the astronomical environments in which vinyl cyanide has been detected, we focus here on collision dynamics in the low-energy regime. Hence, we aim to calculate cross sections for rotationally inelastic transitions up to $E_{\textrm{tot}}=E_{\textrm{k}}+E_{\textrm{rot}} \leq 100$ cm$^{-1}$. The highest rotational levels involved in transitions of vinyl cyanide in the vibrational ground state detected in the ISM involve levels with $j = 10, 11$ \cite{agundez2008detection}. We therefore focus our study on obtaining converged cross sections for transitions involving all levels up to $j = 11$, taking into consideration a sufficient number of open energy channels.\\
The rotational cross sections $\sigma_{j'k'_ak'_c \leftarrow jk_ak_c}$ were computed using the quantum close-coupling (CC) method \cite{arthurs1960theory} for collisions of asymmetric top molecules with helium atoms \cite{green1976rotational,garrison1976coupled} implemented in a parallelized version of \small{MOLSCAT} code \cite{molscat95}. To perform the calculations, the radial coefficients $v_{lm}^{\textrm{sum}}$ were employed in the code. The coupled equations were solved using the diabatic log-derivative propagator \cite{manolopoulos1986improved}. The reduced mass of the interacting system was taken as $\mu = 3.72168$ a.u. (isotopes $^{12}$C, $^{14}$N, $^{1}$H and $^4$He). Cross section convergence tests were performed to select the integration boundaries of the propagator: $R_{\min}$ and $R_{\max}$ were fixed at 3.0 \AA\ and  40 \AA\ respectively. 
We tested the convergence with respect to the size of the rotational basis (defined by the parameter $j_{\max}$) coupled with an energy cutoff (defined by the parameter $E_{\max}$) for several total energies. The results are summarised in Table \ref{tbl:parameters}. The values were selected so that the inelastic cross sections were converged to within 1\%. The possibility of using the coupled-states method to reduce the computational requirements was also investigated, but its use led to errors larger than a factor of two for a large fraction of transitions.\\
\begin{table}
  \caption{$j_{\max}, E_{\max}$ parameters selected for  $E_{\textrm{tot}}$ energy ranges.}
  \label{tbl:parameters}
  \begin{tabular}{ccc}
    \hline
   $E_{\textrm{tot}}$ cm$^{-1}$& $j_{\max}$ & $E_{\max}$ cm$^{-1}$  \\
    \hline
$E_{\textrm{tot}} \leq$ 50 & 16 & 150  \\
50 $< E_{\textrm{tot}} \leq$ 75  & 18 & 200 \\
75 $< E_{\textrm{tot}} \leq$ 100 & 20 & 200 \\
    \hline
  \end{tabular}
\end{table}

\begin{figure}
    \centering
    \includegraphics[width=1.0\linewidth]{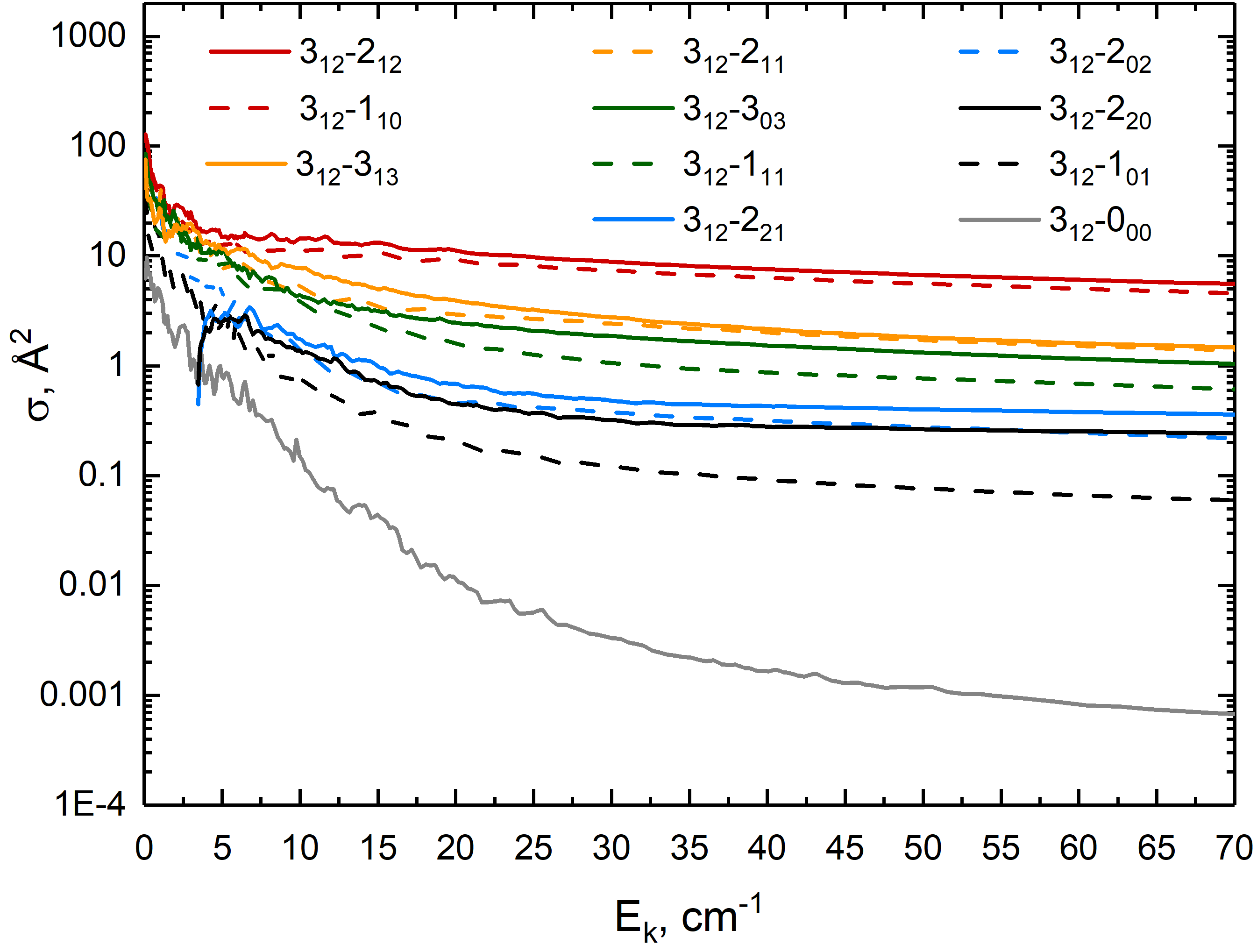}
    \caption{Kinetic energy dependence of the rotational quenching cross-sections $\sigma_{j'k'_ak'_c \leftarrow jk_ak_c}$ of CH$_2$CHCN-He system from the 3$_{12}$ initial state.}
    \label{fig:3_12transitions}
\end{figure}

The cross sections were calculated over a grid of 284 energy points with a variable step size and a larger density of points at low energy to ensure a correct description of the resonance region for the various initial rotational levels. The maximum value of the total angular momentum $J$ was chosen so that the inelastic cross sections were converged within 0.05 \AA$^2$.\\
(De-)excitation cross sections are displayed in Figure \ref{fig:3_12transitions} for the initial rotational level $j_{ka kc}=3_{12}$ with energy $E_{rot}=3.45$ cm$^{-1}$.
The illustrated cross sections all demonstrate similar features at low energies: a dense structure of resonances is observed for $E \leq$ 25 cm$^{-1}$ due to the formation of a quasi-bound collisional complex caused by the presence of the attractive potential well of -53.47 cm$^{-1}$. Outside the resonant region, the cross sections gradually decrease as the collision energy increases.\\
For collision energies above 10 cm$^{-1}$, we observe a clear propensity rule favoring the pair of transitions 3$_{12}$\,--\,2$_{12}$ and 3$_{12}$\,--\,1$_{10}$. This corresponds to transitions with $\Delta j=1$, $\Delta k_a=0$, $\Delta k_c=0$ and $\Delta j=2$ $\Delta, k_a=0$, $\Delta k_c=2$, respectively. The largest cross sections are systematically observed for such transitions involving other initial rotational levels. We also note that both these types of dominating transitions have cross sections that are almost identical over the energy range considered here. The next largest cross sections in Fig. \ref{fig:3_12transitions} are those to levels $3_{13}$ and $2_{11}$, corresponding to transitions $\Delta j= 0,1$ accompanied by $\Delta k_a=0$ and $\Delta k_c=1$.

\begin{figure*}
    \centering
    \includegraphics[width=0.49\linewidth]{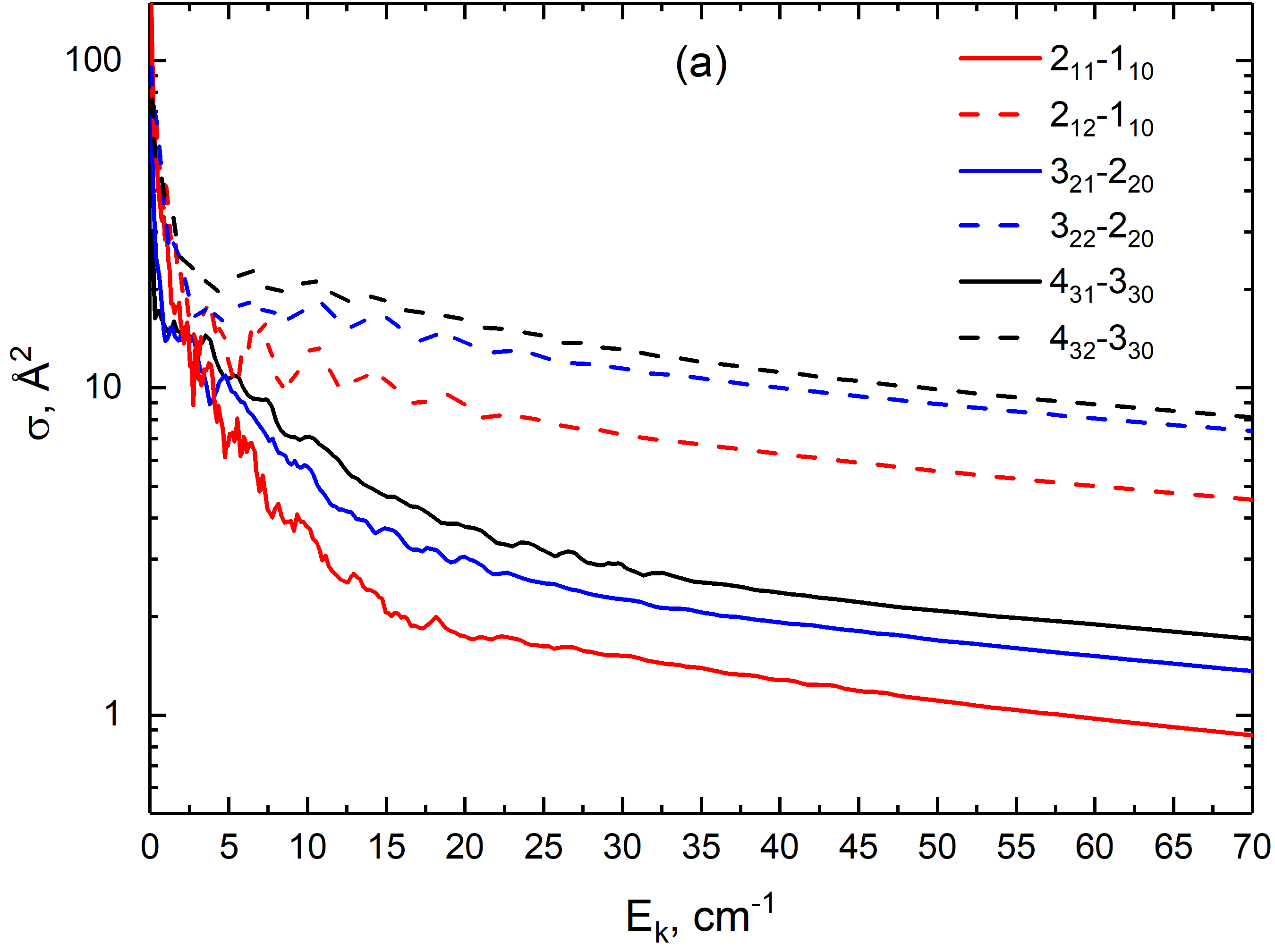}
    \includegraphics[width=0.49\linewidth]{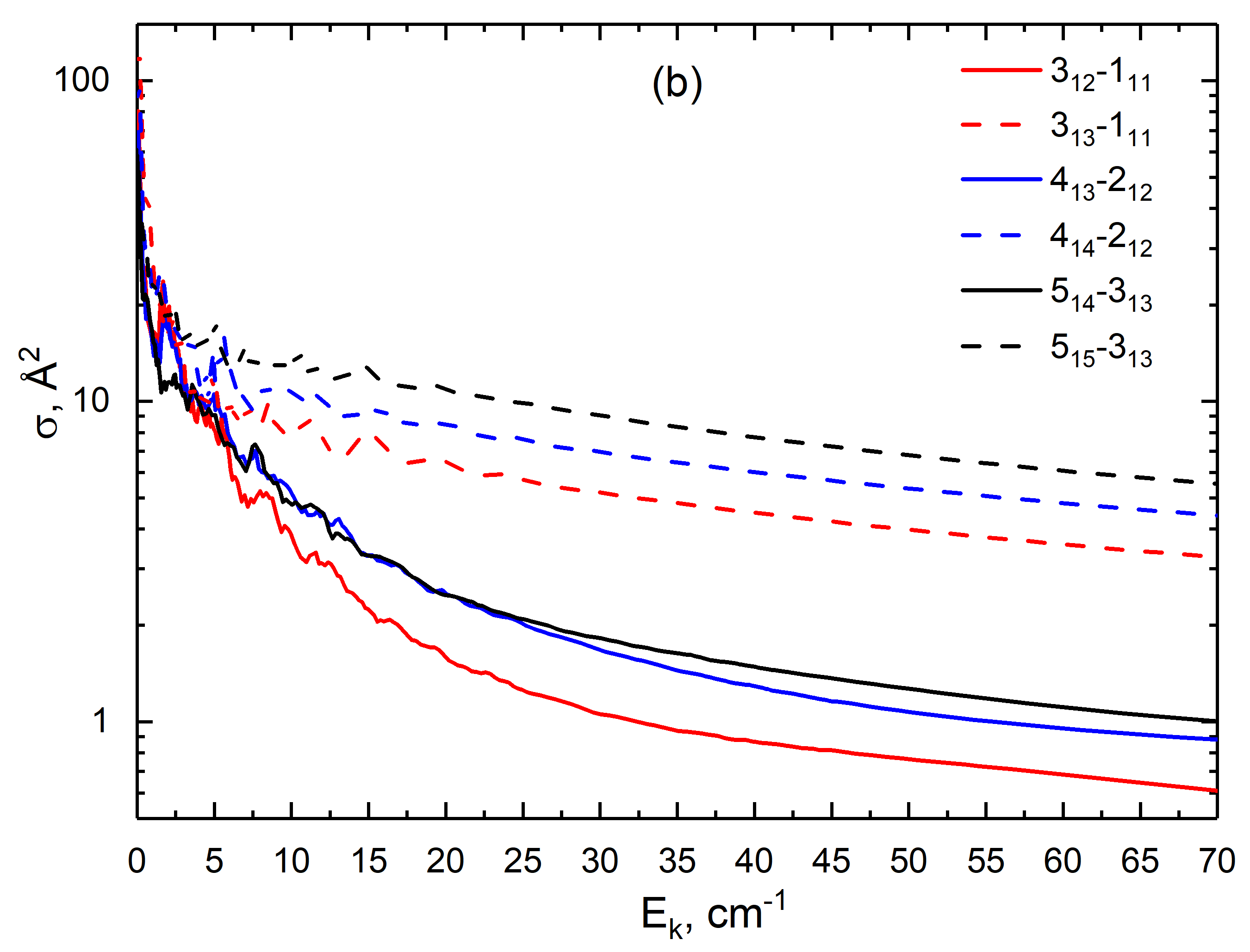}
    \includegraphics[width=0.49\linewidth]{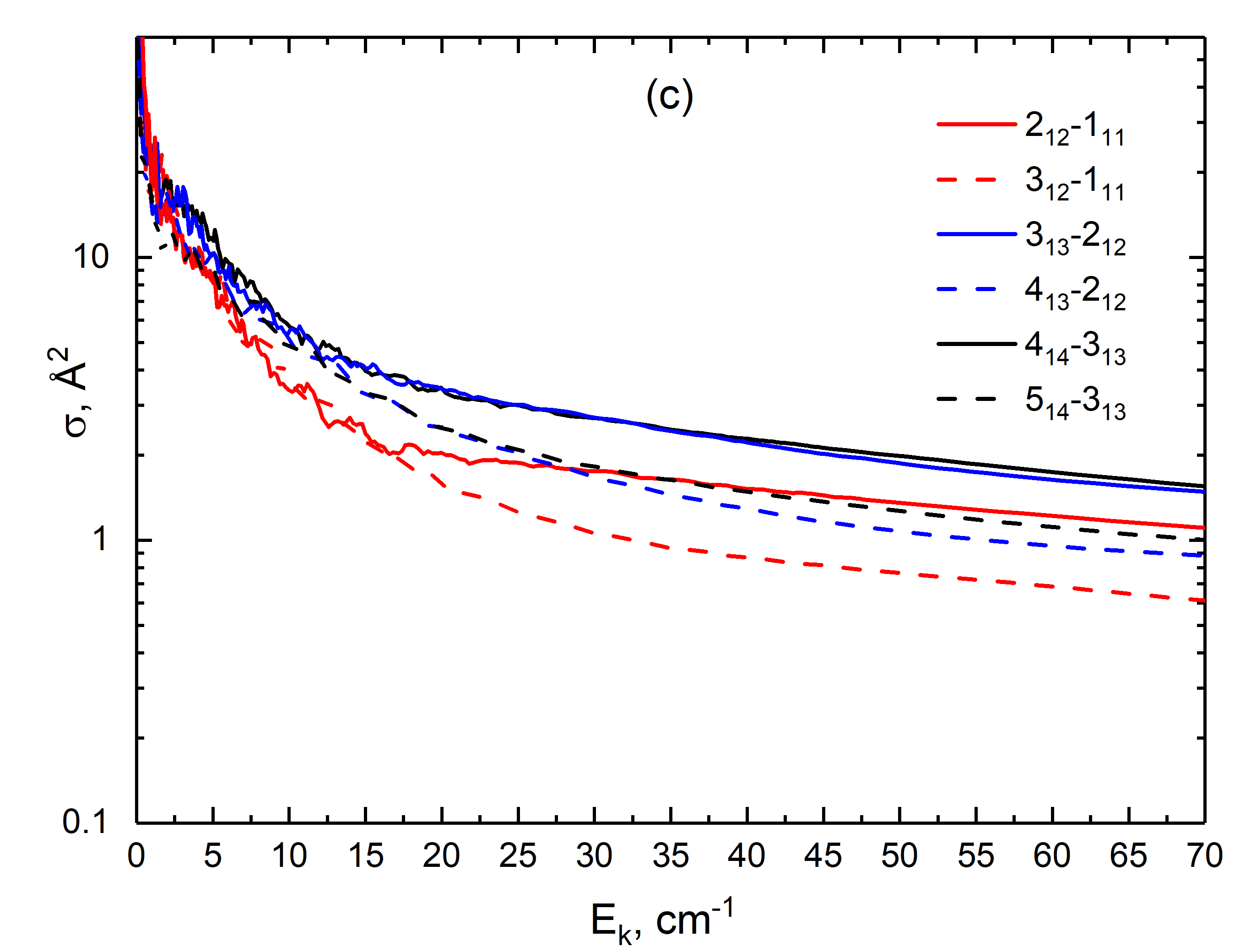}
    \includegraphics[width=0.49\linewidth]{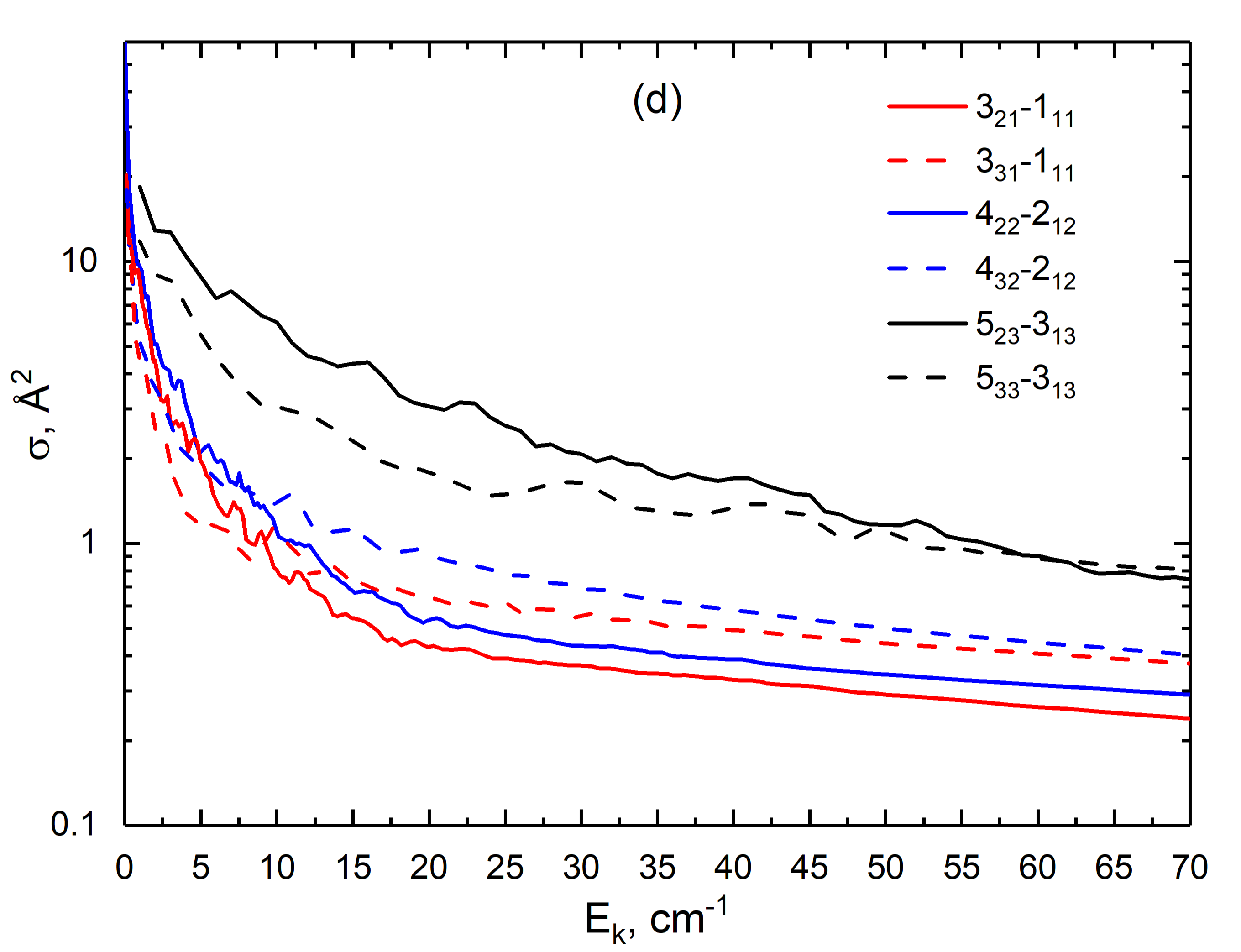}
    \caption{Kinetic energy dependence of the rotational quenching cross sections $\sigma_{j'k'_ak'_c \leftarrow jk_ak_c}$ of CH$_2$CHCN-He system for $\Delta j=1$, $\Delta k_a=0$ and $\Delta k_c=1$ or 2 (panel a); $\Delta j=2$, $\Delta k_a=0$ and $\Delta k_c=1$ or 2 (panel b); $\Delta j=1$ or 2, $\Delta k_a=0$, and $\Delta k_c=1$ (panel c); and for $\Delta j=2$,  $\Delta k_a=1$ or 2, $\Delta k_c=0$ (panel d).}
    \label{crossections}
\end{figure*}

In Figure \ref{crossections} we illustrate more state-to-state rotational quenching cross sections for collisions of CH$_2$CHCN with He atoms as a function of the collisional energy for transitions with specific values of $\Delta j$, $\Delta k_a$ and $\Delta k_c$. Panels a and b display $\Delta j = 1$ and $\Delta j = 2$ transitions with $\Delta k_a=0$ and $\Delta k_c=1$ and 2. Additionally, we also show the state-to-state cross sections for $\Delta j=1$ and 2  $\Delta k_a=0$ and $\Delta k_c=1$ transitions (Panel c) and for $\Delta j=2$  $\Delta k_a=1$ and 2, $\Delta k_c=0$ transitions (Panel d).\\
Figure \ref{crossections} clearly demonstrates that the propensity rules observed for rotational transitions from the initial level $3_{12}$ also hold for other initial states at almost all collision energies. Panels (a) and (b) demonstrate a strong propensity rule favouring transitions with $\Delta k_a=0$ $\Delta k_c=2$. Here, the largest cross sections correspond to the transitions 2$_{11}$\,--\,1$_{10}$, 3$_{22}$\,--\,2$_{20}$, 4$_{32}$\,--\,3$_{30}$ for $\Delta j=1$ and 3$_{13}$\,--\,$1_{11}$, 4$_{14}$\,--\,2$_{12}$, 5$_{15}$\,--\,3$_{13}$ for $\Delta j=2$. Interestingly, the transitions for $\Delta j=1$ and $\Delta j =2$ are similar, although the expansion term $v_{20}$ outweighs the term $v_{10}$, as discussed above.
Other propensity rules are not as prominent, as can be seen on panels (c) and (d), where $\Delta j$ and $\Delta k_a$ propensities are illustrated.
 The clearest trend is thus a propensity rule favoring $\Delta k_a=0$ transitions. This implies a conservation of the angular momentum along the $a$-axis, which corresponds to the smallest moment of inertia. Such propensity rules were already observed for the rotational excitation of other COMs such as propylene oxide in collisions with He atoms \cite{faure2019interaction}.

\subsection{Rate coefficients} \label{rates_section}
From state-to-state cross sections calculated for total energies up to 100 cm$^{-1}$ we computed the corresponding rate coefficients for transitions involving the lowest rotational levels for kinetic temperatures between 5 and 20 K by averaging the cross sections over a Maxwell-Boltzmann distribution:
\begin{equation}
 k_{i \rightarrow f}(T)=\biggl(\frac{8}{\pi\mu\beta}\biggl)^{\frac{1}{2}}\beta^2\int_0^{\infty} E_k \sigma_{i \rightarrow f}(E_k)e^{-\beta E_k} dE_k
\end{equation}
where $\beta$=$1/k_BT$ and $k_B$, $T$ and $\mu$ are the Boltzmann constant, the kinetic temperature and the collision reduced mass respectively.\\
\begin{figure} 
    \centering
    \includegraphics[width=0.9\linewidth]{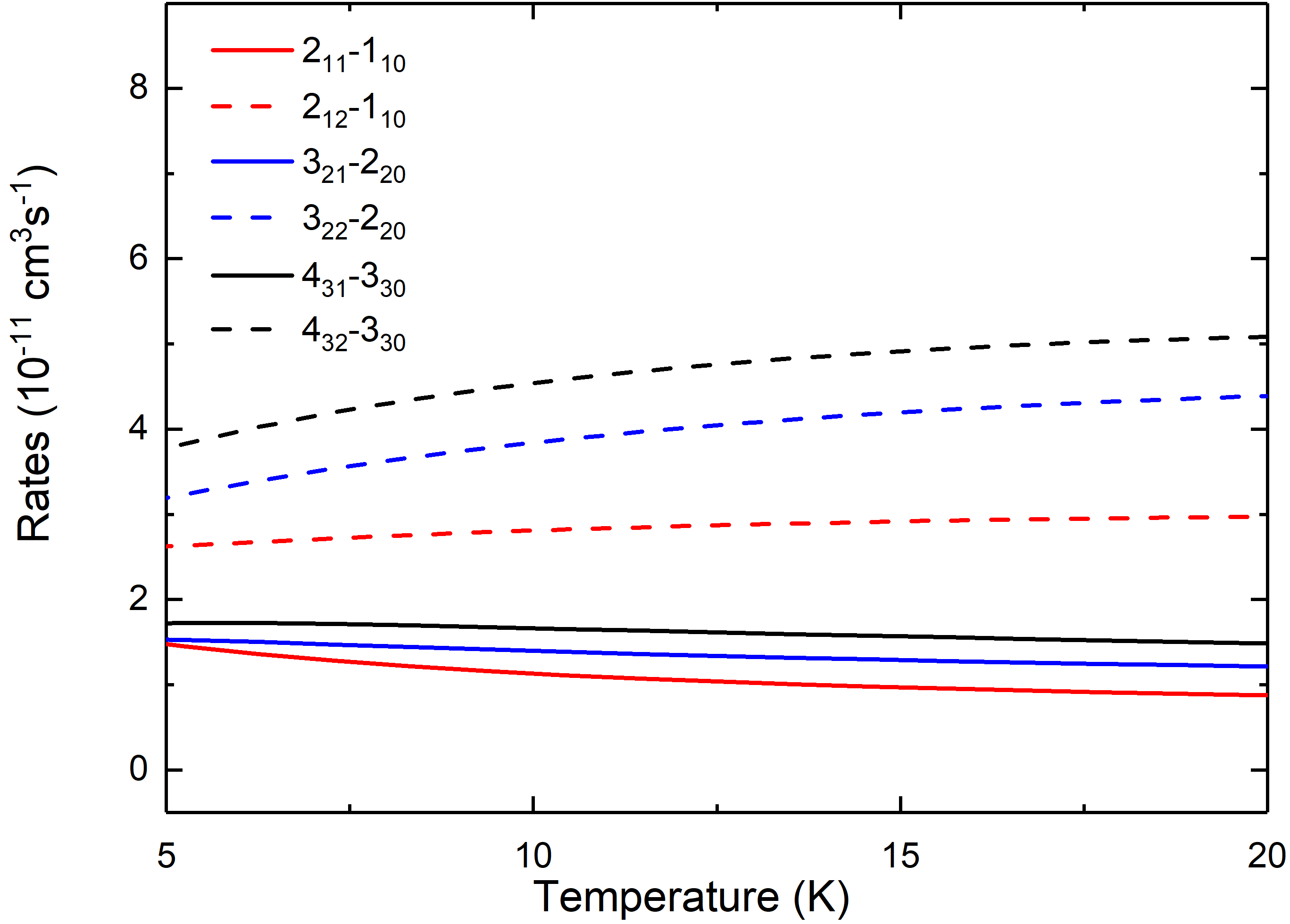}
    \caption{Temperature dependence of the rotational quenching rate coefficients $j'k'_ak'_c \leftarrow jk_ak_c$ of CH$_2$CHCN in collision with He atoms for $\Delta j=1$ $\Delta k_a=0$ and $\Delta k_c=1$ and 2.}
    \label{rate}
\end{figure}
The quenching rate coefficients of CH$_2$CHCN in collision with He are illustrated as a function of kinetic temperature in Figure \ref{rate} for $\Delta j=1$, $\Delta k_a=0$, and $\Delta k_c=1$ or 2 transitions. The rate coefficients do not display a strong dependence on temperature, with a slight increase with temperature for transitions with $\Delta k_c =2$ and a decrease for transitions with $\Delta k_c =1$. In addition, the magnitude of rate coefficients associated with $\Delta k_c=2$ transitions is larger than that of transitions with $\Delta k_c=1$, at all temperatures, reflecting the behaviour of the corresponding cross sections. The other dominant rate coefficients correspond to transitions with $\Delta j=1$, $\Delta k_a=0$, $\Delta k_c=0$, as was observed for cross sections. No other clear propensity rule can be observed.

\section{Conclusions} \label{Conclusions_section}
In this work we developed a strategy to construct the first PES for the CH$_2$CHCN-He system employing a combination of results obtained with two basis sets within the explicitly correlated coupled cluster theory:  CCSD(T)-F12a/aVnZ, n = D,T. This was found necessary due to the large number of \textit{ab initio} points required to obtain an accurate fit of the PES combined with the high cost of these calculations for a large COM such as vinyl cyanide. 
The global minimum of the PES was found at $\phi=51^{\circ}$, $\theta=89^{\circ}$, $R=3.3$ \AA\ with a corresponding potential well depths of $-53.5$ cm$^{-1}$. Several local minima were also found, notably at the geometry $\phi=180^{\circ}$, $\theta=69^{\circ}$, $R=3.6$ \AA, with an energy of $-52.9$ cm$^{-1}$, only 0.6 cm$^{-1}$ above the global minimum. The interaction potential is highly anisotropic in the $\theta$ coordinate with several  local minima separated by barriers. On the other hand, the features of the PES suggest the almost free movement of the He atom along the $\phi$ coordinate.\\
The PES was employed to perform quantum scattering calculations and compute the state-to-state rotationally inelastic cross sections using the quantum-mechanical close-coupling method. Rotational cross sections were calculated as a function of the total energy up to 100 cm$^{-1}$. By thermally averaging the result over the Maxwell--Boltzmann distribution, the state-to-state rotational rate coefficients were obtained for temperatures up to 20 K. Propensity rules favouring transitions with $\Delta k_a=0$ were observed, while other transitions do not demonstrate obvious trends.\\
A further step will be the assessment of the impact of collisional excitation on vinyl cyanide spectra using the obtained rate coefficients in radiative transfer models. This is likely to be important since vinyl cyanide was originally detected as a maser, likely due to the collisional effects. Non-LTE modeling accounting for the competition between the radiative and collisional processes would also allow to better constrain the abundance of these molecules in various astronomical environments and further our understanding of the non-LTE effects. However, this will require extending the present calculations to higher rotational levels and kinetic temperatures.
Finally, we should note that while He is often used as a proxy to model the collisional excitation by para-H$_2$ molecules, this approximation does not necessarily hold \cite{BenKhalifa2021a}, so that studying the excitation of vinyl cyanide by H$_2$ could also be investigated in future work.
\section{Acknowledgments}
JL acknowledges support from KU Leuven through Project no. C14/22/082. MBK acknowledges support from the Research Foundation-Flanders (FWO). The scattering calculations presented in this work were performed on the VSC clusters (Flemish Supercomputer Center), funded by the Research Foundation-Flanders (FWO) and the Flemish Government.

\bibliography{references}

\providecommand{\latin}[1]{#1}
\makeatletter
\providecommand{\doi}
  {\begingroup\let\do\@makeother\dospecials
  \catcode`\{=1 \catcode`\}=2 \doi@aux}
\providecommand{\doi@aux}[1]{\endgroup\texttt{#1}}
\makeatother
\providecommand*\mcitethebibliography{\thebibliography}
\csname @ifundefined\endcsname{endmcitethebibliography}  {\let\endmcitethebibliography\endthebibliography}{}
\begin{mcitethebibliography}{41}
\providecommand*\natexlab[1]{#1}
\providecommand*\mciteSetBstSublistMode[1]{}
\providecommand*\mciteSetBstMaxWidthForm[2]{}
\providecommand*\mciteBstWouldAddEndPuncttrue
  {\def\EndOfBibitem{\unskip.}}
\providecommand*\mciteBstWouldAddEndPunctfalse
  {\let\EndOfBibitem\relax}
\providecommand*\mciteSetBstMidEndSepPunct[3]{}
\providecommand*\mciteSetBstSublistLabelBeginEnd[3]{}
\providecommand*\EndOfBibitem{}
\mciteSetBstSublistMode{f}
\mciteSetBstMaxWidthForm{subitem}{(\alph{mcitesubitemcount})}
\mciteSetBstSublistLabelBeginEnd
  {\mcitemaxwidthsubitemform\space}
  {\relax}
  {\relax}

\bibitem[McGuire(2022)]{mcguire20222021}
McGuire,~B.~A. 2021 census of interstellar, circumstellar, extragalactic, protoplanetary disk, and exoplanetary molecules. \emph{The Astrophysical Journal Supplement Series} \textbf{2022}, \emph{259}, 30\relax
\mciteBstWouldAddEndPuncttrue
\mciteSetBstMidEndSepPunct{\mcitedefaultmidpunct}
{\mcitedefaultendpunct}{\mcitedefaultseppunct}\relax
\EndOfBibitem
\bibitem[Herbst and Van~Dishoeck(2009)Herbst, and Van~Dishoeck]{herbst2009complex}
Herbst,~E.; Van~Dishoeck,~E.~F. Complex organic interstellar molecules. \emph{Annual Review of Astronomy and Astrophysics} \textbf{2009}, \emph{47}, 427--480\relax
\mciteBstWouldAddEndPuncttrue
\mciteSetBstMidEndSepPunct{\mcitedefaultmidpunct}
{\mcitedefaultendpunct}{\mcitedefaultseppunct}\relax
\EndOfBibitem
\bibitem[Yamamoto(2017)]{yamamoto2017introduction}
Yamamoto,~S. Introduction to astrochemistry. \emph{Editorial: Springer} \textbf{2017}, \emph{614}\relax
\mciteBstWouldAddEndPuncttrue
\mciteSetBstMidEndSepPunct{\mcitedefaultmidpunct}
{\mcitedefaultendpunct}{\mcitedefaultseppunct}\relax
\EndOfBibitem
\bibitem[Gardner and Winnewisser(1975)Gardner, and Winnewisser]{gardner1975detection}
Gardner,~F.; Winnewisser,~G. The detection of interstellar vinyl cyanide/acrylonitrile. \emph{Astrophysical Journal, vol. 195, Feb. 1, 1975, pt. 2, p. L127-L130.} \textbf{1975}, \emph{195}, L127--L130\relax
\mciteBstWouldAddEndPuncttrue
\mciteSetBstMidEndSepPunct{\mcitedefaultmidpunct}
{\mcitedefaultendpunct}{\mcitedefaultseppunct}\relax
\EndOfBibitem
\bibitem[Matthews and Sears(1983)Matthews, and Sears]{matthews1983detection}
Matthews,~H.~E.; Sears,~T.~J. The detection of vinyl cyanide in TMC-1. \emph{Astrophysical Journal, Part 1 (ISSN 0004-637X), vol. 272, Sept. 1, 1983, p. 149-153.} \textbf{1983}, \emph{272}, 149--153\relax
\mciteBstWouldAddEndPuncttrue
\mciteSetBstMidEndSepPunct{\mcitedefaultmidpunct}
{\mcitedefaultendpunct}{\mcitedefaultseppunct}\relax
\EndOfBibitem
\bibitem[Ag{\'u}ndez \latin{et~al.}(2008)Ag{\'u}ndez, Fonfr{\'\i}a, Cernicharo, Pardo, and Gu{\'e}lin]{agundez2008detection}
Ag{\'u}ndez,~M.; Fonfr{\'\i}a,~J.~P.; Cernicharo,~J.; Pardo,~J.~R.; Gu{\'e}lin,~M. Detection of circumstellar ch2chcn, ch2cn, ch3cch, and h2cs. \emph{Astronomy \& Astrophysics} \textbf{2008}, \emph{479}, 493--501\relax
\mciteBstWouldAddEndPuncttrue
\mciteSetBstMidEndSepPunct{\mcitedefaultmidpunct}
{\mcitedefaultendpunct}{\mcitedefaultseppunct}\relax
\EndOfBibitem
\bibitem[Balucani(2009)]{balucani2009elementary}
Balucani,~N. Elementary reactions and their role in gas-phase prebiotic chemistry. \emph{International Journal of Molecular Sciences} \textbf{2009}, \emph{10}, 2304--2335\relax
\mciteBstWouldAddEndPuncttrue
\mciteSetBstMidEndSepPunct{\mcitedefaultmidpunct}
{\mcitedefaultendpunct}{\mcitedefaultseppunct}\relax
\EndOfBibitem
\bibitem[Stevenson \latin{et~al.}(2015)Stevenson, Lunine, and Clancy]{stevenson2015membrane}
Stevenson,~J.; Lunine,~J.; Clancy,~P. Membrane alternatives in worlds without oxygen: Creation of an azotosome. \emph{Science advances} \textbf{2015}, \emph{1}, e1400067\relax
\mciteBstWouldAddEndPuncttrue
\mciteSetBstMidEndSepPunct{\mcitedefaultmidpunct}
{\mcitedefaultendpunct}{\mcitedefaultseppunct}\relax
\EndOfBibitem
\bibitem[Palmer \latin{et~al.}(2017)Palmer, Cordiner, Nixon, Charnley, Teanby, Kisiel, Irwin, and Mumma]{palmer2017alma}
Palmer,~M.~Y.; Cordiner,~M.~A.; Nixon,~C.~A.; Charnley,~S.~B.; Teanby,~N.~A.; Kisiel,~Z.; Irwin,~P.~G.; Mumma,~M.~J. ALMA detection and astrobiological potential of vinyl cyanide on Titan. \emph{Science Advances} \textbf{2017}, \emph{3}, e1700022\relax
\mciteBstWouldAddEndPuncttrue
\mciteSetBstMidEndSepPunct{\mcitedefaultmidpunct}
{\mcitedefaultendpunct}{\mcitedefaultseppunct}\relax
\EndOfBibitem
\bibitem[L{\'o}pez \latin{et~al.}(2014)L{\'o}pez, Tercero, Kisiel, Daly, Berm{\'u}dez, Calcutt, Marcelino, Viti, Drouin, Medvedev, \latin{et~al.} others]{lopez2014laboratory}
L{\'o}pez,~A.; Tercero,~B.; Kisiel,~Z.; Daly,~A.~M.; Berm{\'u}dez,~C.; Calcutt,~H.; Marcelino,~N.; Viti,~S.; Drouin,~B.; Medvedev,~I.; others Laboratory characterization and astrophysical detection of vibrationally excited states of vinyl cyanide in Orion-KL. \emph{Astronomy \& Astrophysics} \textbf{2014}, \emph{572}, A44\relax
\mciteBstWouldAddEndPuncttrue
\mciteSetBstMidEndSepPunct{\mcitedefaultmidpunct}
{\mcitedefaultendpunct}{\mcitedefaultseppunct}\relax
\EndOfBibitem
\bibitem[Wakelam \latin{et~al.}(2010)Wakelam, Smith, Herbst, Troe, Geppert, Linnartz, {\"O}berg, Roueff, Ag{\'u}ndez, Pernot, \latin{et~al.} others]{wakelam2010reaction}
Wakelam,~V.; Smith,~I.; Herbst,~E.; Troe,~J.; Geppert,~W.; Linnartz,~H.; {\"O}berg,~K.; Roueff,~E.; Ag{\'u}ndez,~M.; Pernot,~P.~e.; others Reaction networks for interstellar chemical modelling: improvements and challenges. \emph{Space science reviews} \textbf{2010}, \emph{156}, 13--72\relax
\mciteBstWouldAddEndPuncttrue
\mciteSetBstMidEndSepPunct{\mcitedefaultmidpunct}
{\mcitedefaultendpunct}{\mcitedefaultseppunct}\relax
\EndOfBibitem
\bibitem[Balucani \latin{et~al.}(2000)Balucani, Asvany, Chang, Lin, Lee, Kaiser, and Osamura]{balucani2000crossed}
Balucani,~N.; Asvany,~O.; Chang,~A.; Lin,~S.; Lee,~Y.; Kaiser,~R.; Osamura,~Y. Crossed beam reaction of cyano radicals with hydrocarbon molecules. III. Chemical dynamics of vinylcyanide (C 2 H 3 CN; X 1 A′) formation from reaction of CN (X 2 $\Sigma$+) with ethylene, C 2 H 4 (X 1 A g). \emph{The Journal of Chemical Physics} \textbf{2000}, \emph{113}, 8643--8655\relax
\mciteBstWouldAddEndPuncttrue
\mciteSetBstMidEndSepPunct{\mcitedefaultmidpunct}
{\mcitedefaultendpunct}{\mcitedefaultseppunct}\relax
\EndOfBibitem
\bibitem[Vereecken \latin{et~al.}(2003)Vereecken, De~Groof, and Peeters]{vereecken2003temperature}
Vereecken,~L.; De~Groof,~P.; Peeters,~J. Temperature and pressure dependent product distribution of the addition of CN radicals to C 2 H 4. \emph{Physical Chemistry Chemical Physics} \textbf{2003}, \emph{5}, 5070--5076\relax
\mciteBstWouldAddEndPuncttrue
\mciteSetBstMidEndSepPunct{\mcitedefaultmidpunct}
{\mcitedefaultendpunct}{\mcitedefaultseppunct}\relax
\EndOfBibitem
\bibitem[Marchione \latin{et~al.}(2022)Marchione, Mancini, Liang, Vanuzzo, Pirani, Skouteris, Rosi, Casavecchia, and Balucani]{marchione2022unsaturated}
Marchione,~D.; Mancini,~L.; Liang,~P.; Vanuzzo,~G.; Pirani,~F.; Skouteris,~D.; Rosi,~M.; Casavecchia,~P.; Balucani,~N. Unsaturated dinitriles formation routes in extraterrestrial environments: A combined experimental and theoretical investigation of the reaction between cyano radicals and cyanoethene (C2H3CN). \emph{The Journal of Physical Chemistry A} \textbf{2022}, \emph{126}, 3569--3582\relax
\mciteBstWouldAddEndPuncttrue
\mciteSetBstMidEndSepPunct{\mcitedefaultmidpunct}
{\mcitedefaultendpunct}{\mcitedefaultseppunct}\relax
\EndOfBibitem
\bibitem[Dagdigian(2024)]{dagdigian2024rotational}
Dagdigian,~P.~J. Rotational excitation of methanol in collisions with molecular hydrogen. \emph{Monthly Notices of the Royal Astronomical Society} \textbf{2024}, \emph{527}, 2209--2213\relax
\mciteBstWouldAddEndPuncttrue
\mciteSetBstMidEndSepPunct{\mcitedefaultmidpunct}
{\mcitedefaultendpunct}{\mcitedefaultseppunct}\relax
\EndOfBibitem
\bibitem[Walker \latin{et~al.}(2017)Walker, Lique, Dumouchel, and Dawes]{walker2017inelastic}
Walker,~K.~M.; Lique,~F.; Dumouchel,~F.; Dawes,~R. Inelastic rate coefficients for collisions of C6H- with H2 and He. \emph{Monthly Notices of the Royal Astronomical Society} \textbf{2017}, \emph{466}, 831--837\relax
\mciteBstWouldAddEndPuncttrue
\mciteSetBstMidEndSepPunct{\mcitedefaultmidpunct}
{\mcitedefaultendpunct}{\mcitedefaultseppunct}\relax
\EndOfBibitem
\bibitem[Khalifa \latin{et~al.}(2019)Khalifa, Wiesenfeld, and Hammami]{khalifa2019interaction}
Khalifa,~M.~B.; Wiesenfeld,~L.; Hammami,~K. Interaction of the simple carbene cC 3 H 2 with H 2: potential energy surface and low-energy scattering. \emph{Physical Chemistry Chemical Physics} \textbf{2019}, \emph{21}, 9996--10002\relax
\mciteBstWouldAddEndPuncttrue
\mciteSetBstMidEndSepPunct{\mcitedefaultmidpunct}
{\mcitedefaultendpunct}{\mcitedefaultseppunct}\relax
\EndOfBibitem
\bibitem[Ben~Khalifa \latin{et~al.}(2024)Ben~Khalifa, Wiesenfeld, and Loreau]{BenKhalifa2024}
Ben~Khalifa,~M.; Wiesenfeld,~L.; Loreau,~J. Rotational (de-)excitation of CH3CN in collisions with H2 on an accurate potential energy surface (submitted). \emph{Physical Chemistry Chemical Physics} \textbf{2024}, \relax
\mciteBstWouldAddEndPunctfalse
\mciteSetBstMidEndSepPunct{\mcitedefaultmidpunct}
{}{\mcitedefaultseppunct}\relax
\EndOfBibitem
\bibitem[Ben~Khalifa \latin{et~al.}(2022)Ben~Khalifa, Dagdigian, and Loreau]{doi:10.1021/acs.jpca.2c06925}
Ben~Khalifa,~M.; Dagdigian,~P.~J.; Loreau,~J. Interaction of CH3CN and CH3NC with He: Potential Energy Surfaces and Low-Energy Scattering. \emph{The Journal of Physical Chemistry A} \textbf{2022}, \emph{126}, 9658--9666, PMID: 36534035\relax
\mciteBstWouldAddEndPuncttrue
\mciteSetBstMidEndSepPunct{\mcitedefaultmidpunct}
{\mcitedefaultendpunct}{\mcitedefaultseppunct}\relax
\EndOfBibitem
\bibitem[Faure \latin{et~al.}(2019)Faure, Dagdigian, Rist, Dawes, Quintas-Sánchez, Lique, and Hochlaf]{faure2019interaction}
Faure,~A.; Dagdigian,~P.~J.; Rist,~C.; Dawes,~R.; Quintas-Sánchez,~E.; Lique,~F.; Hochlaf,~M. Interaction of chiral propylene oxide (CH3CHCH2O) with helium: potential energy surface and scattering calculations. \emph{ACS Earth and Space Chemistry} \textbf{2019}, \emph{3}, 964--972\relax
\mciteBstWouldAddEndPuncttrue
\mciteSetBstMidEndSepPunct{\mcitedefaultmidpunct}
{\mcitedefaultendpunct}{\mcitedefaultseppunct}\relax
\EndOfBibitem
\bibitem[Ben~Khalifa and Loreau(2023)Ben~Khalifa, and Loreau]{10.1093/mnras/stad3201}
Ben~Khalifa,~M.; Loreau,~J. {Rotational excitation of interstellar benzonitrile by helium atoms}. \emph{Monthly Notices of the Royal Astronomical Society} \textbf{2023}, \emph{527}, 846--854\relax
\mciteBstWouldAddEndPuncttrue
\mciteSetBstMidEndSepPunct{\mcitedefaultmidpunct}
{\mcitedefaultendpunct}{\mcitedefaultseppunct}\relax
\EndOfBibitem
\bibitem[Demes \latin{et~al.}(2024)Demes, Bop, Khalifa, and Lique]{demes2024first}
Demes,~S.; Bop,~C.~T.; Khalifa,~M.~B.; Lique,~F. First close-coupling study of the excitation of a large cyclic molecule: collision of cC 5 H 6 with He. \emph{Physical Chemistry Chemical Physics} \textbf{2024}, \emph{26}, 16829--16837\relax
\mciteBstWouldAddEndPuncttrue
\mciteSetBstMidEndSepPunct{\mcitedefaultmidpunct}
{\mcitedefaultendpunct}{\mcitedefaultseppunct}\relax
\EndOfBibitem
\bibitem[Motte-Tollet \latin{et~al.}(1995)Motte-Tollet, Messina, and Hubin-Franskin]{motte1995electronic}
Motte-Tollet,~F.; Messina,~D.; Hubin-Franskin,~M.-J. Electronic and vibrational excitation of acrylonitrile by low and intermediate energy electrons. \emph{The Journal of chemical physics} \textbf{1995}, \emph{103}, 80--89\relax
\mciteBstWouldAddEndPuncttrue
\mciteSetBstMidEndSepPunct{\mcitedefaultmidpunct}
{\mcitedefaultendpunct}{\mcitedefaultseppunct}\relax
\EndOfBibitem
\bibitem[Buck \latin{et~al.}(1976)Buck, Callomon, Hellwege, Starck, Hirota, Hellwege, Kuchitsu, Lafferty, Maki, and Pote]{buck1976structure}
Buck,~I.; Callomon,~J.; Hellwege,~K.; Starck,~B.; Hirota,~E.; Hellwege,~A.; Kuchitsu,~K.; Lafferty,~W.; Maki,~A.; Pote,~C. \emph{Structure Data of Free Polyatomic Molecules / Strukturdaten freier mehratomiger Molekeln}; Landolt-B{\"o}rnstein: Numerical Data and Functional Relationships in Science and Technology - New Series; Springer Berlin Heidelberg, 1976\relax
\mciteBstWouldAddEndPuncttrue
\mciteSetBstMidEndSepPunct{\mcitedefaultmidpunct}
{\mcitedefaultendpunct}{\mcitedefaultseppunct}\relax
\EndOfBibitem
\bibitem[Werner \latin{et~al.}(2020)Werner, Knowles, Manby, Black, Doll, Heßelmann, Kats, Köhn, Korona, Kreplin, Ma, Miller, Mitrushchenkov, Peterson, Polyak, Rauhut, and Sibaev]{10.1063/5.0005081}
Werner,~H.-J. \latin{et~al.}  {The Molpro quantum chemistry package}. \emph{The Journal of Chemical Physics} \textbf{2020}, \emph{152}, 144107\relax
\mciteBstWouldAddEndPuncttrue
\mciteSetBstMidEndSepPunct{\mcitedefaultmidpunct}
{\mcitedefaultendpunct}{\mcitedefaultseppunct}\relax
\EndOfBibitem
\bibitem[Boys and Bernardi(1970)Boys, and Bernardi]{boys1970calculation}
Boys,~S.~F.; Bernardi,~F. The calculation of small molecular interactions by the differences of separate total energies. Some procedures with reduced errors. \emph{Molecular physics} \textbf{1970}, \emph{19}, 553--566\relax
\mciteBstWouldAddEndPuncttrue
\mciteSetBstMidEndSepPunct{\mcitedefaultmidpunct}
{\mcitedefaultendpunct}{\mcitedefaultseppunct}\relax
\EndOfBibitem
\bibitem[Deegan and Knowles(1994)Deegan, and Knowles]{DEEGAN1994321}
Deegan,~M.~J.; Knowles,~P.~J. Perturbative corrections to account for triple excitations in closed and open shell coupled cluster theories. \emph{Chemical Physics Letters} \textbf{1994}, \emph{227}, 321--326\relax
\mciteBstWouldAddEndPuncttrue
\mciteSetBstMidEndSepPunct{\mcitedefaultmidpunct}
{\mcitedefaultendpunct}{\mcitedefaultseppunct}\relax
\EndOfBibitem
\bibitem[Adler \latin{et~al.}(2007)Adler, Knizia, and Werner]{adler2007simple}
Adler,~T.~B.; Knizia,~G.; Werner,~H.-J. A simple and efficient CCSD (T)-F12 approximation. \emph{The Journal of chemical physics} \textbf{2007}, \emph{127}\relax
\mciteBstWouldAddEndPuncttrue
\mciteSetBstMidEndSepPunct{\mcitedefaultmidpunct}
{\mcitedefaultendpunct}{\mcitedefaultseppunct}\relax
\EndOfBibitem
\bibitem[Dunning(1989)]{10.1063/1.456153}
Dunning,~J.,~Thom~H. {Gaussian basis sets for use in correlated molecular calculations. I. The atoms boron through neon and hydrogen}. \emph{The Journal of Chemical Physics} \textbf{1989}, \emph{90}, 1007--1023\relax
\mciteBstWouldAddEndPuncttrue
\mciteSetBstMidEndSepPunct{\mcitedefaultmidpunct}
{\mcitedefaultendpunct}{\mcitedefaultseppunct}\relax
\EndOfBibitem
\bibitem[Knizia \latin{et~al.}(2009)Knizia, Adler, and Werner]{knizia2009simplified}
Knizia,~G.; Adler,~T.~B.; Werner,~H.-J. Simplified CCSD (T)-F12 methods: Theory and benchmarks. \emph{The Journal of chemical physics} \textbf{2009}, \emph{130}\relax
\mciteBstWouldAddEndPuncttrue
\mciteSetBstMidEndSepPunct{\mcitedefaultmidpunct}
{\mcitedefaultendpunct}{\mcitedefaultseppunct}\relax
\EndOfBibitem
\bibitem[van~der Avoird \latin{et~al.}(1980)van~der Avoird, Wormer, Mulder, and Berns]{van1980ab}
van~der Avoird,~A.; Wormer,~P.~E.; Mulder,~F.; Berns,~R.~M. Ab initio studies of the interactions in Van der Waals molecules. \emph{Van der Waals Systems} \textbf{1980}, 1--51\relax
\mciteBstWouldAddEndPuncttrue
\mciteSetBstMidEndSepPunct{\mcitedefaultmidpunct}
{\mcitedefaultendpunct}{\mcitedefaultseppunct}\relax
\EndOfBibitem
\bibitem[Schmitt and Meerts(2018)Schmitt, and Meerts]{schmitt2018structures}
Schmitt,~M.; Meerts,~L. \emph{Frontiers and advances in molecular spectroscopy}; Elsevier, 2018; pp 143--193\relax
\mciteBstWouldAddEndPuncttrue
\mciteSetBstMidEndSepPunct{\mcitedefaultmidpunct}
{\mcitedefaultendpunct}{\mcitedefaultseppunct}\relax
\EndOfBibitem
\bibitem[Flower(2007)]{flower2007molecular}
Flower,~D. \emph{Molecular collisions in the interstellar medium}; Cambridge University Press, 2007; Vol.~42\relax
\mciteBstWouldAddEndPuncttrue
\mciteSetBstMidEndSepPunct{\mcitedefaultmidpunct}
{\mcitedefaultendpunct}{\mcitedefaultseppunct}\relax
\EndOfBibitem
\bibitem[Johnson(1999)]{147901}
Johnson,~R. NIST 101. Computational Chemistry Comparison and Benchmark Database. 1999\relax
\mciteBstWouldAddEndPuncttrue
\mciteSetBstMidEndSepPunct{\mcitedefaultmidpunct}
{\mcitedefaultendpunct}{\mcitedefaultseppunct}\relax
\EndOfBibitem
\bibitem[Arthurs and Dalgarno(1960)Arthurs, and Dalgarno]{arthurs1960theory}
Arthurs,~A.; Dalgarno,~A. The theory of scattering by a rigid rotator. \emph{Proceedings of the Royal Society of London. Series A. Mathematical and Physical Sciences} \textbf{1960}, \emph{256}, 540--551\relax
\mciteBstWouldAddEndPuncttrue
\mciteSetBstMidEndSepPunct{\mcitedefaultmidpunct}
{\mcitedefaultendpunct}{\mcitedefaultseppunct}\relax
\EndOfBibitem
\bibitem[Green(1976)]{green1976rotational}
Green,~S. Rotational excitation of symmetric top molecules by collisions with atoms: Close coupling, coupled states, and effective potential calculations for NH3--He. \emph{The Journal of Chemical Physics} \textbf{1976}, \emph{64}, 3463--3473\relax
\mciteBstWouldAddEndPuncttrue
\mciteSetBstMidEndSepPunct{\mcitedefaultmidpunct}
{\mcitedefaultendpunct}{\mcitedefaultseppunct}\relax
\EndOfBibitem
\bibitem[Garrison \latin{et~al.}(1976)Garrison, Lester, and Miller]{garrison1976coupled}
Garrison,~B.~J.; Lester,~W.~A.; Miller,~W.~H. Coupled-channel study of rotational excitation of a rigid asymmetric top by atom impact:(H2CO, He) at interstellar temperatures. \emph{The Journal of Chemical Physics} \textbf{1976}, \emph{65}, 2193--2200\relax
\mciteBstWouldAddEndPuncttrue
\mciteSetBstMidEndSepPunct{\mcitedefaultmidpunct}
{\mcitedefaultendpunct}{\mcitedefaultseppunct}\relax
\EndOfBibitem
\bibitem[{Hutson} and {Green}(1995){Hutson}, and {Green}]{molscat95}
{Hutson},~J.~M.; {Green},~S. MOLSCAT computer code, version 14, Distributed by Collaborative Computational Project 6. Warington, UK: Daresbury Laboratory, 1995\relax
\mciteBstWouldAddEndPuncttrue
\mciteSetBstMidEndSepPunct{\mcitedefaultmidpunct}
{\mcitedefaultendpunct}{\mcitedefaultseppunct}\relax
\EndOfBibitem
\bibitem[Manolopoulos(1986)]{manolopoulos1986improved}
Manolopoulos,~D. An improved log derivative method for inelastic scattering. \emph{The Journal of chemical physics} \textbf{1986}, \emph{85}, 6425--6429\relax
\mciteBstWouldAddEndPuncttrue
\mciteSetBstMidEndSepPunct{\mcitedefaultmidpunct}
{\mcitedefaultendpunct}{\mcitedefaultseppunct}\relax
\EndOfBibitem
\bibitem[Ben~Khalifa and Loreau(2021)Ben~Khalifa, and Loreau]{BenKhalifa2021a}
Ben~Khalifa,~M.; Loreau,~J. {Fine and hyperfine excitation of nitric oxide by collision with para-H$_2$ at low temperature}. \emph{Mon. Not. R. Astron. Soc.} \textbf{2021}, \emph{508}, 1908--1914\relax
\mciteBstWouldAddEndPuncttrue
\mciteSetBstMidEndSepPunct{\mcitedefaultmidpunct}
{\mcitedefaultendpunct}{\mcitedefaultseppunct}\relax
\EndOfBibitem
\end{mcitethebibliography}

\end{document}